\documentclass{aastex}
\usepackage{emulateapj5,graphicx,epsfig,times}
\usepackage{onecolfloat}  

\shorttitle{Universality of Cluster $(U-V)$ CMRs}
\shortauthors{McIntosh et al.}
\slugcomment{{\sc Accepted for publication in the Astrophysical Journal: } October 2, 2004}

\newcommand\rp{$R_{\rm p}$ }
\newcommand\hst{{\it HST}}

\newcommand\kms{km~s$^{-1}$}
\newcommand\hkpc{$h^{-1}$~kpc }              
\newcommand\hmpc{$h^{-1}$~Mpc }              
\newcommand\rhl{$r_{\rm hl}$ }

\newcommand\cmrsig{$\sigma_{\rm CMR}$ }

\begin{document}

\def\head{

\title{Testing the Universality of the $(U-V)$ Color-Magnitude Relations for Nearby Clusters of Galaxies}

\author{Daniel H.\ McIntosh$^{\rm a,b}$, Ann I.\ Zabludoff$^{\rm a}$, Hans-Walter Rix$^{\rm c}$, Nelson Caldwell$^{\rm d}$}
\affil{$^{\rm a}$Steward Observatory, University of Arizona, 933 North Cherry Avenue, Tucson, AZ 85721}
\affil{$^{\rm b}$present address: University of Massachusetts, Lederle Graduate Research Tower, Amherst, MA 01003; \texttt{dmac@hamerkop.astro.umass.edu}}
\affil{$^{\rm c}$Max-Planck-Institut f\"{u}r Astronomie, Heidelberg, Germany}
\affil{$^{\rm d}$Smithsonian Astrophysical Observatory, Cambridge, MA 02138}
\vspace{0.2cm}

\begin{abstract}
We present a detailed $(U-V)$ color-magnitude relation (CMR) analysis for
three local ($z<0.06$) clusters of galaxies.  
From square-degree imaging of the Abell clusters A85, A496, and A754, 
we select 637 galaxies down to $L\sim0.1L^{\star}_V$ with
spectroscopic membership, which minimizes
uncertain field contamination corrections.  To characterize the degree of
CMR uniformity among nearby clusters, we use a maximum likelihood technique
to quantify the CMR properties of the red galaxies in each cluster.
We find that these clusters have similar CMRs with a mean
color of $(U-V)=1.40$ at $M_V=-20+5\log_{10}h$, and narrow limits
of intrinsic color scatter [0.047,0.079] and slope [-0.094,-0.075].
If we restrict our analysis to the core cluster population of red galaxies,
the resultant CMRs are in close 
agreement with that of the Coma Cluster, the only local cluster with a
rest-frame $(U-V)$ CMR determination of comparable precision.  
Therefore, the CMR uniformity of present-day clusters spans a fairly
wide range of cluster masses.
We test how sensitive the CMR uniformity is to variations in color aperture 
size and sample selection, and we find at most slightly wider limits
of scatter [0.047,0.112] and slope [-0.104,-0.054].  This
upper limit for $(U-V)$ scatter is consistent with the bulk of the
stellar populations of red cluster galaxies forming before $z=1.2$, with
a maximum age spread of 5.2 Gyr.  In addition, we note that using colors
from apertures containing equal fractions of galactic light does not
remove the CMR slope and that none of the slopes exhibit
a break as claimed by Metcalf et al.  Our findings expand the
single Coma data point and provide a much-needed
$z=0$ baseline for comparisons to high redshift cluster CMRs at the same
rest-frame wavelengths.  The range in CMR scatter that we find
among nearby clusters is consistent with the values reported for
clusters at higher redshifts, further suggesting that there has been little
evolution in the stellar populations of red-sequence cluster galaxies out to at
least $z\sim1$.
To identify the most recently accreted cluster galaxies, 
we divide each cluster's membership into three galaxy
populations based on $(U-V)$ color relative to the well-defined CMR.
Blue and moderately blue galaxies make up
$18-23\%$ by number of each cluster population more luminous
than $0.1L^{\star}$.
Our color-magnitude division should represent a rough time since 
cluster accretion.
In testing this hypothesis, we find that blue galaxies are spatially,
kinematically, and morphologically distinct from red cluster galaxies.
Even in projection, the blue galaxies reside towards 
the outskirts of the cluster and appear to avoid the inner half megaparsec, 
in contrast with the increasing density of red sequence galaxies towards
the cluster center.  In addition, the blue galaxies have
velocity distributions relative to the cluster rest frame that are flatter,
and some have different means, compared to the roughly Gaussian distribution 
of red member velocities.
Members with the bluest colors tend to be disk-like or irregular
in appearance compared to the red galaxies, which
have mostly early-type (E/S0)
morphologies.  Moderately blue cluster galaxies may be an intermediate mix 
with a fraction of small bulge-to-disk ratio S0s, yet these require closer
scrutiny.  The spatial, kinematic, and morphological distinctions
between blue and red cluster galaxies provide further evidence that 
CMR-relative color is related to time since cluster infall, and that
bluer members are indeed the most recently
accreted field spirals as expected in an hierarchical universe.
\end{abstract}

\keywords{cosmology: observations --- galaxies: clusters: general ---
galaxies: clusters: individual (A85,A496,A754) --- galaxies: elliptical and
lenticular, cD --- galaxies: evolution ---
galaxies: photometry}
} 

\twocolumn[\head]  

\section {Introduction}

The color-magnitude (C-M) diagrams 
of rich galaxy clusters provide information essential to understanding
the formation and evolution of cluster members.  For example,
the early-type (E/S0) galaxies in clusters
follow a well-defined correlation of increasingly red galaxy color with
growing luminosity \citep[e.g.][]{visvanathan77,bower92b,hogg04}, 
which is often labeled the color-magnitude relation (CMR) or the 
``red sequence''.  The apparent uniformity of the slope  
and the small color scatter of the CMR are interpreted conventionally as
evidence that these galaxies have ancient stellar populations
($z_{\rm f} \sim 2-3$) with a small range in relative ages
\citep{bower92b,lubin96,kodama98,bower98}.
Furthermore, if the CMR is well-determined, it is possible to identify
the population of most recently accreted cluster members by their
blue colors relative to the CMR \citep{balogh00,kodama01a,bicker02}.
Such recent arrivals to the cluster
environment make ideal tests of the picture of hierarchical cluster
formation and the mechanisms that may affect the evolution of cluster
galaxies.  

Because of its particular sensitivity to recent star formation (SF),
the near-UV C-M diagram (i.e., $U-V$ color vs. $V$-band
absolute magnitude) has the potential to place the best constraints on the  
formation epoch and relative ages of the red sequence galaxies, as
well as on the properties of the infalling population of cluster
galaxies.  Unfortunately, due to the difficulty in obtaining 
wide-field imaging and spectroscopic follow-up for large samples of
cluster members, little is known about the near-UV properties of
nearby cluster galaxies.  In this paper, we present the results from a
new wide-field spectroscopic and near-UV imaging study of galaxy clusters
in the nearby ($z<0.06$) universe.

Constraining the era over which the stars in the oldest cluster
galaxies formed is a subject of much study and debate 
\citep[][and references therein]{bower98,gladders98,rose01,kodama01a,vandokkum01a}.  The
near-UV CMRs of nearby clusters have an important
role to play in addressing this question.  Because $(U-V)$ color
straddles the 4000 \AA\ break, near-UV CMRs are more sensitive than red
CMRs to any small differences in the star-formation histories (SFHs) of  
cluster galaxies.  For example, simple stellar burst models
\citep[e.g.][]{worthey94,vazdekis96} show that a stellar population with
solar metallicity that formed $>10$ Gyr ago over a period of 4 Gyr
has an intrinsic color scatter in $(U-V)$ that is roughly twice as large as
in $(B-R)$, regardless of the model details (i.e. initial mass function,
metallicity, etc.).  If the near-UV CMRs of nearby clusters are uniform 
in slope and narrow in scatter, then the formation time and age
spread for the stars can be better constrained than with $(B-R)$.
Another constraint on
this formation epoch can be obtained by comparing the near-UV CMRs of nearby
clusters to the CMRs of distant clusters based
on observations at red optical passbands, which
correspond to the same rest-frame near-UV band.  Recent work on
distant clusters has shown that the CMR scatter
based on rest-frame $(U-V)$ colors is quite small and
homogeneous for redshifts $z>0.5$ \citep{ellis97,stanford98}.
The observed absence of evolution in the scatter of the near-UV CMR 
at moderate redshifts suggests an early formation time for the
constituent stars, but a direct comparison to large, well-defined 
samples of $z\sim0$ cluster galaxies is lacking.

Early work on the near-UV CMRs, which included only the brightest,
early-type members of a few poorly-sampled nearby clusters, suggested
that the near-UV CMRs are universal \citep{visvanathan77,sandage78,bower92b}.
More recent work has raised questions about the form of cluster CMRs.
For example, there is controversy as to whether the slope of
the relation changes at lower luminosities \citep{metcalfe94}.
To date, Coma (A1656, $z=0.022$) remains the only rich cluster at low
redshift with a well-defined near-UV CMR based on hundreds of
spectroscopically-confirmed cluster members sampled from large
cluster-centric distances \citep[$>1$~Mpc;][]{terlevich01}.
As a consequence, it is hard to assess the degree
of uniformity among cluster CMRs and to compare them with high
redshift systems.  Determining the near-UV CMRs for a larger sample of
clusters as we have now done is therefore essential to resolve these
issues.

Identifying the population of galaxies most recently accreted by the
cluster is critical to (1) test how lower mass systems like galaxies
and groups combine to form clusters and (2) determine the importance of
factors that may influence the evolution of cluster galaxies, e.g.
ram pressure stripping \citep{abadi99,quilis00},
gas consumption through starbursts 
\citep{hashimoto98,bekki98,fujita99,rose01} or
strangulation \citep{larson80,balogh00b,bekki02},
galaxy harassment \citep{moore98,moore99}, and galaxy-galaxy interactions
or mergers in infalling groups \citep{zabludoff96,zabludoff98}.
Compared to the population of old, early types that define
the CMR, spectroscopic studies of galaxies with bluer colors at each
luminosity show them to have younger luminosity-weighted mean ages
\citep[e.g.][]{vandokkum98,terlevich99}.  
Therefore, it follows that the bluest cluster members relative to the CMR
have SF properties most similar to the field population of spirals, and
thereby are likely to be the most recent cluster arrivals
\citep{balogh00,kodama01a,bicker02}.
This population has only been identified in clusters at intermediate redshifts
\citep[$z>0.3$][]{dressler97,vandokkum98,vandokkum99,poggianti99,kodama01c}.
In nearby clusters, many of the infalling galaxies may lie at large
projected radii \rp from the cluster center \citep{diaferio01}, making them
difficult to observe with small area detectors.  For example, 
nearby clusters appear
to have small ($<5\%$) fractions of blue galaxies \citep{bo78a,bo84},
which is likely the result of an observational bias if new members are
expected at large $R_{\rm p}$.  Until quite recently only the
cores ($R_{\rm p}<0.5$ $h^{-1}$~Mpc) of local ($z<0.1$)
clusters had been imaged due to the limiting size of detectors and the large
area these objects span on the sky.  A detailed search for the most recent cluster
arrivals in Abell clusters at $z<0.06$ requires wide-field imaging
($\ge0.5$ degree per 1 $h^{-1}$~Mpc).   Here we utilize a wide-field
(one square degree) imaging campaign to acquire data for a large number
of cluster galaxies out to $R_{\rm p}>1$~\hmpc. 
With our wide-field spectroscopic sample of nearby
clusters, it is possible to ask whether the
spatial, kinematic, and morphological properties of the bluest cluster
galaxies support the picture that they are new cluster members.
 
This paper is part of a series based on a complete sample of 637 bright 
galaxies with precise $(U-V)$ color data and spectroscopic redshifts
confirming their residence in
three nearby Abell clusters: A85 ($z=0.055$, R=1, $\sigma=993$~\kms), 
A496 ($z=0.033$, R=1, $\sigma=728$~\kms) and 
A754 ($z=0.055$, R=2, $\sigma=953$~\kms); cluster richness R and
velocity dispersion $\sigma$ are described in Tables~1
and 7, respectively.  We tabulate the general properties of the 
three clusters in Table~1.  This large catalog of cluster galaxy
photometry is complete to several magnitudes below $L^{\star}$\footnote{We 
assume $M^{\star}_V = -20.6 + 5\log_{10}h$ for red-sequence galaxies
from $M^{\star}_B = -19.7 + 5\log_{10}h$ \citep{binney98} and a mean
E/S0 galaxy color of $(B-V)=0.90$~mag \citep{fukugita95}.}, extends
well outside of each parent cluster core, and effectively quadruples
the previous sample of clusters with similarly rich $(U-V)$ data on 
spectroscopic members.
The clusters were selected according to the following criteria:
(a) observable from the northern hemisphere ($\delta > -15\arcdeg$);
(b) out of the Galactic plane ($|b| > 15\arcdeg$);
(c) close enough to resolve recent structural and morphological changes
in galaxies with ground-based images from a 1-meter class
telescope ($<0.5$~\hkpc~pix$^{-1}$),
yet far enough to be in relatively smooth Hubble
flow beyond the Local Supercluster ($0.03\leq z \leq 0.07$); and (d)
each having
$>100$ spectroscopic members from the literature.
At the time of our observations the three clusters selected had the largest
number of available redshifts among local Abell clusters, 
roughly 1500 total from \citet{christlein03}.  Spectroscopic membership provides
the advantage of not requiring background contamination corrections,
which is especially important at large cluster radii.

In our first paper \citep[][hereafter Paper~1]{mcintosh04}
we present evidence for environment-driven galaxy transformation
through a detailed and quantitative comparison of cluster members and field
galaxies with similar luminosities and blue colors.  In
Paper~3 (D. H. McIntosh, H.-W. Rix, \& N. Caldwell, in preparation)
we explore the structure and quantitative morphology of this sample of
cluster members
as a function of luminosity, cluster density, and cluster-centric radius.
The outline of this paper is as follows.  We provide
the details of our observations, data reduction, calibration, sample selection,
and galaxy photometry in \S2.  In \S3.1 we analyze the cluster
CMRs by measuring their properties (slope, scatter, and zero point) using
a maximum-likelihood fitting technique.  We quantify the degree of
uniformity among local cluster CMRs in \S3.2.  In \S3.3 we identify the
most recent populations of infalling galaxies through their position
in C-M space relative to each cluster's CMR, and we explore their
spatial, kinematic, and morphological properties.  We present our
conclusions in \S4.
Throughout this paper we use $h = H_0/(100$~km~s$^{-1}$~Mpc$^{-1})$, and we
assume a $\Lambda$-CDM (cold dark matter),
$\Omega_{\rm M}=0.3$ and $\Omega_{\rm \Lambda}=0.7$, flat ($\Omega_{\rm k}=0$)
cosmology.  

\section{Observations and Data}

\subsection{Cluster Imaging}
We observed clusters A85, A496, and A754 during two runs at 
the Kitt Peak National
Observatory (KPNO) 0.9-meter Telescope with the NOAO Mosaic Wide-Field
Imager \citep{boroson94,muller98}.  We used the engineering-grade
Mosaic for the $V$ data in November 1997, and the science-grade Mosaic
with thinned CCDs for the $U$-band in January 2000.  We imaged each cluster
in each passband
cluster with a minimum of five exposures dithered a few arcminutes
on the sky for the removal of inter-chip gaps, cosmic rays and other defects.
The dither pattern we used ensured at least $80\%$
of the maximal exposure
for all regions of a final combined image from five exposures.
The initial exposure of a dithered set was centered on the brightest cluster
galaxy's position given in Table~1.  The Mosaic imagers
provide a $59\arcmin \times 59\arcmin$ field of view with an
unbinned pixel scale of $0\farcs423$ (15 $\micron$ pixels).  This
wide-field coverage spans physical scales of 1.6 (A496) and 2.7\hmpc
(A85 and A754).  
We had at least one photometric night per each Mosaic run during
which we observed hundreds of \citet{landolt92} standard stars for
calibrating our cluster galaxy data to \citet{johnson66} $U$ and $V$ 
magnitudes (see \S \ref{PhotSys}).  Our cluster observations are
summarized in Table~2.

The KPNO 0.9-meter telescope is a Richey-Chretien design with an $f/3.3$ Cer-Vit
primary and an $f/7.5$ secondary.  This $f/7.5$ system produces a curved
focal plane; therefore, a 2-element
fused silica field corrector is required for wide-field imaging with the
Mosaic camera.  This imager employs
large (5.75 inches square) par-focal $U$ and $V$ filters.

\begin{figure}
\includegraphics[scale=.95, angle=0]{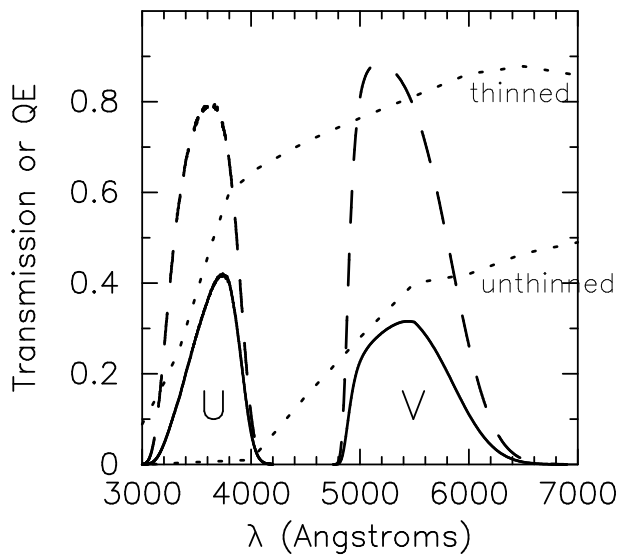}
\caption{Average Mosaic response in the $U$ and $V$ passbands (solid lines).
The dotted
lines represent the average quantum efficiency (QE)
of the eight Mosaic Loral CCDs for
the thinned, science-grade ($U$-band, MOS2) and unthinned, engineering-grade
($V$-band, MOS1) cases.  The transmission of each filter is given by the dashed
lines.  The total response is the combination of the filter transmission
and QE.  The $U$-band response of our observations drops off
rapidly blueward of $\sim3750$~\AA, similar to the Coma observations of 
\protect\citet{bower92a}.
\label{FiltResp} }
\vspace{-0.2cm}
\end{figure}

The Mosaic imaging CCD system has eight $2048$\,$\times$\,$4096$
pixel Loral chips.  The $V$-band data were taken with the original
unthinned, engineering-grade chips (hereafter MOS1), which result in lower
quantum efficiency (QE) as shown in
Figure~\ref{FiltResp}.  For these observations the readout noise,
dark current, and average single chip gain were $12.25 e^-$, 
$\sim100$~to~$250  e^-$~hour$^{-1}$, and $6.64 e^-$~ADU$^{-1}$, respectively.
The $U$-band data were obtained following an upgrade \citep{wolfe98}
in which the original CCDs were replaced with
thinned, science-grade (AR coated SITe) chips (hereafter MOS2).
Even with the improved sensitivity the QE of the science-grade detector
falls off rapidly blueward of 4000\AA\ (Figure~\ref{FiltResp});
therefore, the spectral response in the $U$-band is not a perfect match
to standard \citet{johnson66} $U$-band, which was obtained with a blue
sensitive photomultiplier.  Nevertheless,
the use of photometric standards with a limited range of $(U-V)$ colors
allows calibration to a standard photometric
system (as described in \S \ref{PhotSys}).
The read noise, dark current, and average single chip gain for 
the $U$-band observations were
$5.66 e^-$, $\sim15 e^-$~hour$^{-1}$, and
$2.86  e^-$~ADU$^{-1}$, respectively.

\subsection{Data Reduction}
\label{Redux}
For each cluster in our study we want to construct deep integration, 
wide-field $U$ and $V$ images
devoid of bad pixels, cosmic rays, and gaps between CCDs.
Achieving such high quality images requires
stacking dithered Mosaic frames, which places high demands on the initial data
reduction steps\footnote{NOAO CCD Mosaic Imager User Manual (hereafter
MosManual), Version Sept. 15, 2000,
G. Jacoby; http://www.noao.edu/kpno/mosaic/manual/index.html.}.
In particular, the data must be well-flattened and carefully
corrected for photometric effects of the variable pixel scale.
To this end we employ
a customized reduction pipeline that uses the
IRAF\footnote{IRAF is
distributed by the National Optical Astronomical Observatories, which are
operated by AURA, Inc. under contract to the NSF.}
environment and adheres to standard image reduction techniques.

We perform basic reduction of each individual Mosaic frame using the IRAF
{\sc mscred} package.  This software allows image processing to be
performed on a multi-chip exposure as if it was a single CCD frame.
For each single Mosaic exposure, we fit a first order Legendre polynomial
to the overscan regions of the individual chip images, and then use the fits
to correct (``debias'') each chip image.  Following debiasing we trim the
chip images of their overscan regions.  We include a small correction
($<0.1\%$ MOS1, $<0.3\%$ MOS2) that is necessary due to cross talk between
pairs of adjacent chips sharing the same electronics in
the Mosaic detectors; i.e., for each $(i,j)$ pixel of chips \#1,3,6,8 a
predetermined fraction of the $(i,j)^{\rm th}$ pixel value from its
corresponding adjacent chip (i.e. \#2,4,5,7)
is subtracted.  We then subtract the average bias (using seven zero exposures)
level from each Mosaic frame.  The $V$-band
images from MOS1 require a dark correction due to
significant dark current ($2-5$ ADU per chip per exposure).
Roughly $3.1\%$ ($0.4\%$) of the MOS1 (MOS2) pixels are bad.  We flag
bad pixels and include them in an image mask
used during the final image combination.
In addition, a large portion ($1000\times1000$~pixel) of the upper
northeast corner of MOS1 chip\#5 appears to be adversely affected by many
hot pixels with variable dark current.  We remove this region as we cannot
correct for it adequately.  We note that the removal of this bad region
represents only $1.6\%$ of the
total area covered by the $V$-band imaging, resulting in the loss of 
roughly 2-4 galaxies per cluster at large cluster-centric distance.  This
loss does not affect our results.

An important step towards achieving precise photometry is the determination and
removal of the response function of the individual CCDs -- i.e.
flat fielding the data.
Traditionally, dividing each exposure by a uniformally
illuminated blank frame will produce an image that has a uniform and
flat appearance.  Yet,
the Mosaic imagers have pixel scales that decrease roughly quadratically
such that an individual pixel in a field corner is 6\% smaller,
and contains only 92\% of the flux, compared to a pixel at field center
(see MosManual for details).  Therefore, although an individual star anywhere
on the image will have the same number of photons within the
point-spread-function (PSF), the variable pixel scale causes the photometric
zero point to vary by 8\% over the field of view.  We correct for this
photometric effect following the recommendations given in the MosManual.
Briefly, we flatten each image with a flat-field frame that has {\it not}
been corrected for the variable pixel scale.  Then, following our astrometric
calibration and prior to stacking multiple exposures, we re-grid each
frame to a tangent-plane projection with pixels of constant angular scale.
We note that during the re-gridding, we {\it do not} scale each
pixel photometrically by the amount each pixel area has changed.  In
this manner we account for the variable pixel scale and produce
uniform images over the entire field of view.

We construct flat-field frames for each passband using a combination of
twilight and night-sky flats.  First we make a normalized flat with high
S/N by averaging a set of twilight illuminated exposures.
This component accounts for both the small scale, high frequency
(pixel-to-pixel) variations in response, and the spatial variations over
large fractions of an image.  We then fit
a smooth surface to a night-sky flat produced by median combining a set of
unflattened cluster frames with all objects masked out, and we multiply
this smooth surface by the twilight flat frame to produce a high S/N flat
with the same spectral response as our science observations.
We iterate the flat field construction twice to optimize the night-sky flat
object masking.  Thus, the resultant ``super flat'' for each passband is
spatially and spectrally flat.
We divide each cluster and standard star image by its
super flat to achieve $\lesssim1\%$ deviations from global flatness 
over the eight chip array.

Good astrometry is required to register and stack each dithered set of
exposures.  In addition, accurate celestial coordinates
are necessary to identify galaxies with measured redshifts.
All Mosaic images have an initial default world coordinate system (WCS) loaded
in their header at the time of observation.  The WCS maps the image pixel
space onto celestial coordinate axes (RA and Dec).  However, effects such as
global pointing offsets, instrument rotations, and differential atmospheric
refraction produce the need for corrections to this astrometric calibration.
We use the {\sc msccmatch} package in IRAF to interactively
derive accurate (RMS~$\lesssim 0.3\arcsec$) astrometry for each
dithered exposure by registering $\sim300$ stars
(with $12.0<V<16.5$) distributed evenly over each image to their epoch
J2000.0 equatorial coordinates in the USNO-v2.0 system \citep{monet96}.
We then remap the eight chip exposures for
each Mosaic frame onto a single image by rebinning
the pixels to a tangent-plane projection, thus producing pixels of constant
angular size (as described above).
At this point our standard star frames are fully reduced.

Before combining a dithered sequence of registered exposures into a single
high S/N image, we must account for the
effects of differing transparency between individual frames.  Transparency
differences are due to time varying effects during a dither sequence.
The most common effect is the changing airmass over an hour long dithered
exposure.  Other effects, resulting in variable photometric conditions
(e.g. thin cirrus, clouds) can change the atmospheric transparency.
First we subtract a constant flux level equal to the mode sky value
from each image in a dithered set.  Higher order terms are unnecessary due
to the better than 1\% global flat fielding.  Next all
exposures of a dithered sequence are scaled to a reference
image with the lowest
airmass and/or best photometric conditions.
We calculate the multiplicative scale factor
for each image by comparing simple aperture flux measurements for the set of
$\sim300$ astrometric calibration stars common between each image and its
reference image.  Scaling factors have typical values within a few percent
of unity.
Therefore, the set of precombined cluster frames are scaled
to the same effective airmass and exposure time.
We note that variable airmass conditions can affect color, but we find
the relative airmass and $(U-V)$ color dependence to be
negligible (see \S~\ref{PhotSys}).

Finally, we median combine each dithered set of scaled and
sky-subtracted cluster images.  We use median combination, rather than
a cosmic ray rejection algorithm, to produce final images free of most cosmic 
rays.  We restore the sky level to the average background calculated from the 
modes of each image multiplied by their corresponding transparency scale
factors.  In Table~3
we tabulate the relevant properties of the final cluster images.

\subsection{Photometric System}
\label{PhotSys}
We transform our cluster galaxy photometry to the \citet{landolt92}
system.  We selected photometric flux standards with colors similar to
the range of colors observed in typical cluster galaxies at low redshift
$0.0\leq (U-V) \leq2.1$, and which have been observed more than twice
by \citet{landolt92} to ensure photometric consistency.  
During nights with apparently photometric
conditions we observed $\sim100$~to~$200$ photometric standard stars in each
passband.
To achieve good photometric calibration,
we observed standard fields at least three times during the night at
relative airmass readings
in the range $1.2\leq X \leq2.0$.

The photometric solution is based on a simple, linear transformation equation
that relates standard magnitudes to instrumental magnitudes in each band pass.
The total flux $f_{\rm ap}$ from an exposure of duration $t_{\rm exp}$
within a circular aperture of $14\farcs4$
diameter,
similar to that used by \citet{landolt92}, is converted to an instrumental
magnitude
\begin{equation}
m_{\rm instr} = -2.5 (\log_{10}{f_{\rm ap}} - \log_{10}{t_{\rm exp}}) .
\label{InsMag}
\end{equation}
Then we use IRAF's {\sc photcal} package on the selected
set of standard star $m_{\rm instr}$ measurements to find the best-fit
solutions to the following transformation equations:
\begin{equation}
V_{\rm L92} = m_{\rm instr} + zp_V + \alpha_V X_V + \beta_V (U-V)_{\rm L92}
\label{VTrnsfEqn}
\end{equation}
\begin{equation}
U_{\rm L92} = m_{\rm instr} + zp_U + \alpha_U X_U + \beta_U (U-V)_{\rm L92} .
\label{UTrnsfEqn}
\end{equation}
The photometric system is defined by the coefficients ($zp$, $\alpha$ and
$\beta$) of the transformation
equation in each passband.  The photometric zero point $zp$ quantifies
the gain and
the total sensitivity of the telescope plus detector.  The airmass term $\alpha$
is a measure of the atmospheric extinction as a function of telescope altitude.
The
color term $\beta$ shows how well the instrumental system matches the
\citet{landolt92} system.  A color dependence on airmass is known to exist
especially at bluer wavelengths; however,
through testing we find no need for an
airmass-color cross term in the transformation equation for our photometric
calibrations.

The observed standards primarily fall on the central chips (i.e. \#2,3,6,7);
however, recall that we obtain the instrumental magnitudes from
fully reduced images that have been corrected to have a globally uniform
zero point (see \S \ref{Redux}).  We solve the
corresponding transformation equation for each nightly set of standards in
a given passband and present the
coefficients and their formal errors in Table~4.
We note that the apparent systematic difference
of 0.10 mag between the two $V$-band calibration zero points in fact
amounts only to a 0.02 mag difference when the airmass terms are included.
The photometric accuracy ($\sim3\%$ error in $zp$) of this data is close
to optimum given the remapping of the eight chips.  In this case,
the dominant
source of the photometric uncertainty is due to CCD-to-CCD sensitivity
differences (MosManual).

With the above photometric system and a flux measurement $f$, the
magnitude for each galaxy is
\begin{equation}
m = zp_{\rm eff} - 2.5\log_{10}{f} ,
\label{mageq}
\end{equation}
where the effective zero point 
$zp_{\rm eff} = zp + \alpha \tilde{X} + 2.5\log_{10}{\tilde{t}_{\rm exp}}$
accounts for the effective exposure time and airmass of each final
image as given in Table~3.

\subsection {Selection of Cluster Members}

\subsubsection{Source Detection and Deblending}
Given the large numbers of sources in each square-degree image, we use the
source detection and extraction software
SExtractor \citep{bertin96} to automatically construct catalogs 
of accurate source positions from the $U$ and $V$ final frames.
In addition, SExtractor provides instrumental
magnitudes \\ (MAG\_BEST) and other photometric parameters for each detected
source.  We configure SExtractor to
detect objects comprised of a minimum of 5 connected pixels (DEBLEND\_MINAREA)
above a background threshold of $3\sigma_{\rm bkg}$ (DETECT\_THRESH).
Overlapping sources are deblended into multiple objects if the contrast  
between flux peaks associated with each object is $\geq0.05$ (DEBLEND\_MINCONT).
These parameters provide our
working definition of an imaged source.  We confirm that these parameters
provide good source detection and deblending by visually inspecting
random regions from each image.  We reject sources flagged (FLAGS$\ge4$)
as saturated, or otherwise bad, either of which will
corrupt the deblending and extraction routines of SExtractor.
In addition, we exclude sources within 140 pixels (1 arcmin) of image
edges; these regions are lower in S/N due to the dithering of the
individual frames. 
For each catalog we determine an empirical magnitude 
limit $m_{\rm min}$ where the source counts distribution
flattens and begins to fall off.  We summarize the magnitude limits and 
number counts for our final $U$ and $V$ cluster source catalogs in 
Table~5.

\subsubsection{Star/Galaxy Separation}
\label{stargal}
Owing to the variable PSF as a function of image position, stellar profiles
can be extended especially towards the image edges.  Moreover, small
circular galaxies are unresolved especially at faint magnitudes.  
Therefore, we do not rely on the SExtractor star/galaxy classifier
(CLASS\_STAR) to separate stars and galaxies.  Instead, we use
GIM2D \citep{simard02}
to fit a PSF-convolved, bulge$+$disk model to the
two-dimensional (2D) surface brightness profile of all sources in the $V$-band
catalog.  We select the $V$-band catalog due to its higher S/N and better
seeing characteristics.  GIM2D provides the model half-light 
(or effective) radius \rhl and the total flux $f_{\rm tot}$, among other
structural parameters.  In Paper~3 we present a detailed description of
our galaxy profile fitting technique and the complete catalog of 
structural properties measured from our cluster data.  

In this paper we use
the fluxes to calculate total $V$-band magnitudes and the half-light size
to separate stars and galaxies.  
Though stars are not fit well by a two component
model, the PSF convolution produces $r_{\rm hl}<1$~pixel
model profiles for most stars.  In contrast, galaxies brighter than $V=19$
have \rhl significantly larger than the seeing size.
Thus, our method provides robust star/galaxy
separation to $V\leq19$, one magnitude fainter than the limit where
our $U,V$ photometry and redshift matching are quite ($\ge95\%$) complete.
With this method we find a total of 4320 extended sources (i.e. galaxies)
among the three clusters, of which 1469 are brighter than  $V=19$ mag.
We cross-correlate the
$V$-band selected galaxies with the magnitude-limited $U$ sources to
produce a catalog of 1315 total galaxies with $U,V$ photometric 
data and accurate
coordinates from the three cluster fields.  We note that our cross-correlation 
with $U$-band data at similar limiting depth as the $V$-band 
creates a color bias such
that at the faintest magnitudes only increasingly bluer sources are matched.
Therefore, as shown in Figure \ref{zuvCts} 
we find $U$-band detections for 1093 (74\%) of the $V\leq19$
galaxies; while we successfully match 95\% (727/765) brighter than $V=18$ mag.
We give the source counts breakdown for each cluster in Table~5.

\subsubsection{Total Magnitudes}
\label{totmags}
We use GIM2D model fluxes from profile fits to the $V$-band imaging to
calculate the total magnitudes (equation~\ref{mageq}) 
for all $V$-selected extended
sources.  We present the $V_{\rm tot}$ distribution for the combined
cluster (A85$+$A496$+$A754) galaxy sample in Figure~\ref{zuvCts}.
GIM2D estimates parameter uncertainties through full Monte-Carlo
propagation of the parameter probability distributions during each best-fit
model determination.  Typical mean errors in $V_{\rm tot}$ are 0.04 mag,
which is significantly larger than the $V$-band Poisson errors of cluster
members ($\le0.01$ mag).
In Figure~\ref{magcomp} we
compare the GIM2D-based total magnitudes against the SExtractor
(MAG\_BEST)\footnote{SExtractor automatically selects between several
photometric measurements based on degree of influence from neighboring
sources.} derived values for all galaxies with $U$ and $V$ detections.
GIM2D $V_{\rm tot}$ magnitudes are 10\% brighter on average than
SExtractor determinations.  The GIM2D total flux measurement includes
light from the LSB regions at large galactic radii, whereas SExtractor
magnitudes are aperture-based and thus underestimate the total flux
systematically when large isophotal thresholds, such as $3\sigma_{\rm bkg}$,
are used \citep{bertin96}.  We emphasize here that we do not use
SExtractor magnitudes in any of our analysis.  

Throughout this paper we use GIM2D-derived
$V_{\rm tot}$ magnitudes to calculate rest-frame absolute magnitudes
\begin{equation}
M_V - 5\log_{10}h = V_{\rm tot} - DM - A_V + k_V + C ,
\label{TotMagEqn}
\end{equation}
where $DM$ is the cosmological distance modulus if $h=1$, 
$A_V=3.315\times E(B-V)$
is the Galactic extinction correction following 
\citet[][hereafter SFD98]{schlegel98}, $k_V$ is the $k$-correction from
\citet{poggianti97}, and $C$ is a minor photometric color correction
(see \S~\ref{colors}) amounting to roughly $+0.01$ mag on average.
These magnitudes have random uncertainties of $\approx0.02-0.07$ mag, with
the dominant source of random error being the uncertainty in the GIM2D model 
flux.  We present $M_V$ measurements with formal random errors
for a small sample of cluster members
in Table~6; the full catalog is available electronically.
In addition, sources of systematic error in $M_V$ include uncertainty
in the overall photometric zero point (0.02 mag), the relative extinction
zero point (0.07 mag), and $k$-correction (0.01 mag).

We use the mean cluster redshift $z_{\rm clust}$ to
calculate the distance moduli for individual cluster members, which removes the
possibility of introducing magnitude uncertainties due to galaxy motions
through the cluster.  Assuming each cluster is a sphere with a characteristic
radius of $1.5$~\hmpc (consistent with an Abell radius), we estimate a
$\sim2\%$ error in our adopted distance moduli due to the uncertainty of
front-to-back cluster distance.
We present each cluster's mean redshift and resulting cosmological
distance modulus in Table~7.

We correct the galaxy magnitudes for the effects of Galactic
extinction along the line of sight using the dust maps of SFD98, which
provide reddening $E(B-V)$ values as a function of Galactic coordinates $(l,b)$.
The average corrections are 0.12 (A85), 0.42 (A496),
and 0.21 (A754) $V$ magnitudes.
The correction for A496 is particularly large due to the presence of a
Galactic molecular cloud along the line of sight
(Finkbeiner, D.\ 2001, private communication).
SFD98 adopt a systematic uncertainty in extinction correction of 0.020~mag in $E(B-V)$
(i.e. $\sim0.07$ mag in $M_V$), which represents the median difference
between their reddening estimates from $100\micron$ dust maps
and the 21cm gas maps of \citet{burstein82}.  In addition, SFD98 give a formal
uncertainty of $10\%$ in $E(B-V)$, which is 0.01 (A85), 0.04 (A496),
and 0.02 (A754) in $M_V$.

Even at the low redshifts ($z<0.06$) of our cluster galaxies, the observed
shift of the rest-frame spectral energy distributions is a
significant effect, making galaxies appear redder and dimmer, requiring
a $k$-correction. 
We use the elliptical (E-model) $k$-corrections from \citet{poggianti97},
which amount to mean corrections of -0.06 mag (A496) and -0.10 mag (A85,A754)
to the rest-frame $V$-band.
At $z<0.06$, the three basic Poggianti models (E,Sa,Sc) give $k$-corrections
that differ by $<1\%$.  We adopt 0.01~mag from the model-dependent 
difference as
an estimate for the systematic uncertainty in the \citet{poggianti97}
$k$-corrections.

\subsubsection{Confirmed Member Galaxy Catalogs}
\label{confmem}
By analyzing only known cluster members we remove the need for uncertain
foreground/background statistical corrections.  
\citet{christlein03} analyzed 1486 recessional velocities along the line
of sight towards the three clusters we study here.  
These redshifts were obtained from galaxy spectra in
$1.5\times 1.5\arcdeg$ multi-fiber fields with central coordinates similar to
our cluster images; therefore, about 35\% of the spectroscopic data are
outside of our imaging.

We select member galaxies based on mean recessional velocity 
$\left<cz\right>_{\rm clus}$ and internal velocity dispersion
$\sigma_{\rm clus}$ measurements for each cluster from \citet{christlein03}.
We define cluster members as those galaxies with recessional velocities
$cz_i=\left<cz\right>_{\rm clus} \pm 3\sigma_{\rm clus}$, where the cluster
parameters are given in Table~7.
There are 971 galaxies with redshifts that have coordinates within the 
regions bounded by our cluster imaging,
of which 721 are cluster members and 250 fall outside of
our membership definition.  
We cross-correlate the coordinates of galaxies from the redshift data
with our $U,V$ source positions from our imaging catalog and achieve a total
of 793 image/redshift matches (637 members and 156 non-members).
We define image/redshift matches to be the nearest within
$5\arcsec$ and find the mean coordinate separation is $<2\arcsec$.
The image/redshift membership breaks down into
180 (A85), 146 (A496), and 311 (A754) galaxies; we present
the relevant information for each cluster membership in Table~7.
The final sample contains a total of 637 spectroscopic
cluster member galaxies with a large range of absolute magnitudes
($-16.5 < M_V < -23.0$).  The $U,V$ C-M data for
A85, A496 and A754 are comparable in depth and in membership ($N=275$) to
the most comprehensive study of the Coma cluster by \citet{terlevich01}.

Within the region of our imaging, 84 spectroscopic
members have no $U,V$ matched image
counterparts, of which only 12 are brighter than $V=18$ mag and
72 are fainter than this limit.  Therefore, the completeness for our galaxy 
image/redshift cross correlation for cluster members is 97.6\% (494/506)
at $V\leq18$ mag.  Nearly half (5/12) of the incompleteness at $V\le18$
is due to $>5\arcsec$
coordinate mismatch between the spectroscopy and imaging catalogs, while
two more have imaging that is contaminated by nearby saturated stars.  The
remaining five are comprised of two E/S0s and three spirals.  The
fainter ($V>18$) unmatched redshifts are missed for
several reasons:  24\% (17/72) with $\theta_{\rm sep}>5\arcsec$; 7\% (5/72) 
contaminated by nearby saturated stars;
49\% (35/72) are even fainter ($V>19.5$ mag) and/or are spheroidal
in appearance and likely red in color, which in turn drops them out of
our $U$-band limited catalog of $U,V$ matches; and 21\% (15/72) have disk
morphologies in the $V$-band, either LSB or edge-on, in each case also
not detected in our $U$-band imaging.  Additionally, there are 94 non-cluster
redshifts not detected in both $U$ and $V$ frames.
All but five are background sources, thus, we assume these have been
missed due to red $(U-V)$ color making them fainter than our $U$-band cut.

\subsubsection{Sample Completeness}
\begin{figure}[t]
\center{\includegraphics[scale=1, angle=0]{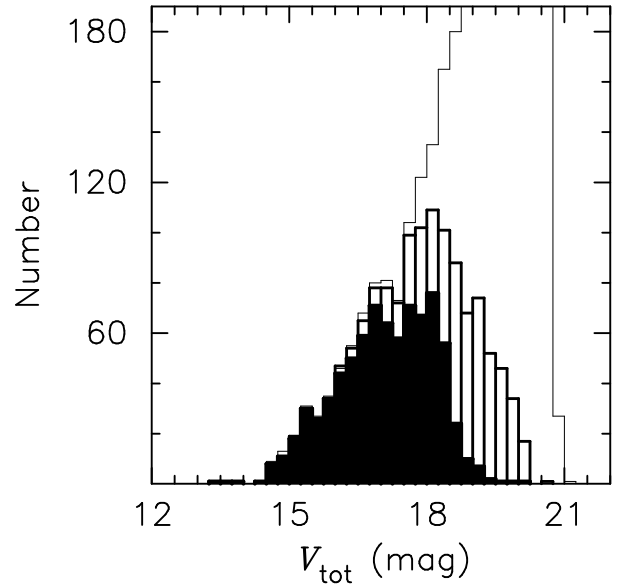}}
\caption{Extended source counts in our cluster images as a function of total
magnitude $V_{\rm tot}$ from GIM2D fits to the $V$-band imaging.  For the
combined cluster sample we
plot the number counts distribution for 4320 extended sources
from our $V$-band limited catalog (solid outline), for 1315 sources
with $U$- and $V$-band photometry (binned histogram), and for 793 galaxies
with spectroscopic redshifts matched to extended source coordinates from
our $U,V$ imaging (solid histogram).  Among the 793 $U,V$ sources with 
redshifts, 637 are cluster members and 156 are non-member, mostly background,
galaxies.  The $V$-band extended source
distribution has been clipped to resolve the level of completeness in the
imaging and spectroscopy.  At $V\le18$ we have $U$-band photometry for
95\% (727/765) of our $V$-band detections, and we find that 85\% 
(615/727) of these have redshifts.  Furthermore, within our imaging
97.6\% (494/506) of cluster member redshifts brighter than $V=18$ have
$U,V$ source counterparts.
\label{zuvCts} }
\vspace{-0.2cm}
\end{figure}

\begin{figure*}
\center{\includegraphics[scale=.8, angle=0]{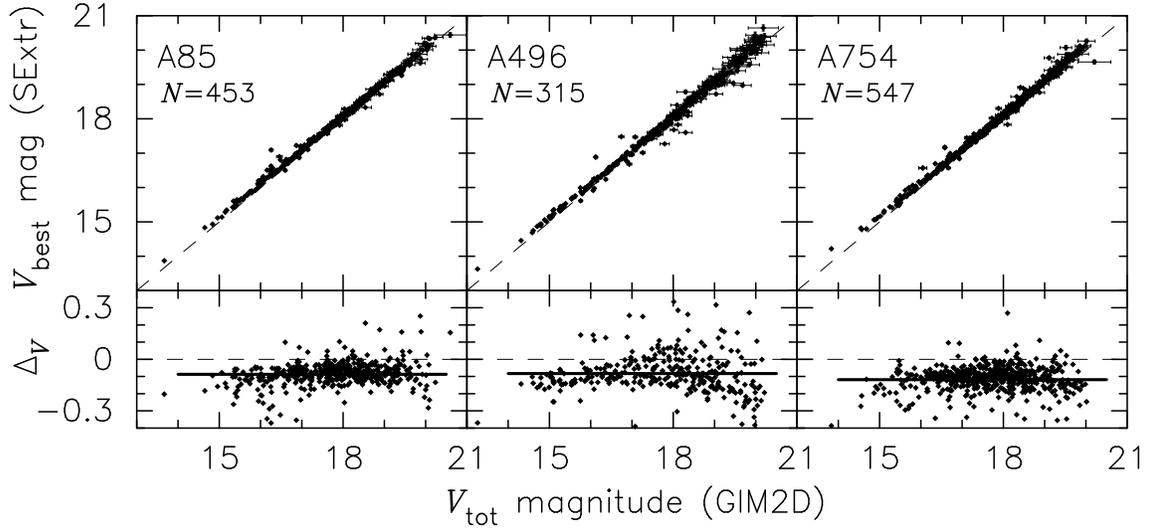}}
\caption{Comparison of GIM2D total $V$-band magnitudes with SExtractor
MAG\_BEST measurements for the complete sample of cluster galaxies with
$U$ and $V$ detections.  The total number of galaxies from each cluster
image are noted in each panel (see Table~5).  The lower
panels show the magnitude difference (GIM2D-SExtractor).
The bold line represents the mean magnitude differences for each cluster:
$-0.087\pm0.004$ (A85), $-0.083\pm0.009$ (A496), and $-0.118\pm0.004$ (A754).
The scatter in $\Delta_V$ for each cluster is, from left to right:
0.08, 0.16, and 0.09 mag.
\label{magcomp} }
\vspace{-0.2cm}
\end{figure*}

For the combined cluster sample in Figure~\ref{zuvCts}, we plot the
distribution of galaxy counts per $V$ magnitude for sources
with spectroscopic identification (members and non-members) in comparison
with our imaging catalog counts for $U,V$ matched and $V$-band extended sources.
This illustrates the completeness of our matched $U$ and $V$ extended
source detections (95.0\% for $V\le18$).  We have shown in the previous
section that for $V\le18$ spectroscopic members we have a 97.6\% complete
sample of $U$ and $V$ image sources.
Finally, we have hundreds of unidentified (i.e. no spectroscopic 
counterparts)
extended sources with $U,V$ photometry towards each cluster.  At $V\le18$
there are 112 imaged extended sources without redshifts. These represent the
expected fractional incompleteness of the spectroscopic sample
\citep{christlein03} and provide a direct measure of the completeness of our
cluster galaxy observations, 84.6\% (615/727) for $V\le18$.

\subsubsection{Morphological Classification}
\label{mmakeup}
Using the $V$-band images we classify cluster members into three coarse visual
types: (1) elliptical, (2) S0, and (3) spiral or irregular.  For the bulk of our
analysis we will concentrate on the subset of 546 members brighter than 
$M_V=-18.1+5\log_{10}h$ ($0.1 L^{\star}$).
This magnitude limit is within 0.1~mag of 
the relatively complete cutoff of $V=18.0$ for A85 and A754,
and represents an extremely complete sample for A496 at $V=16.9$ mag.
We use the isophotal contours of each galaxy to distinguish between ellipticals,
which have smooth radial profiles, and lenticular (S0) galaxies with
separate bulge and disk morphological components characterized by an
intensity discontinuity \citep{dressler80b}.  Most
bright disk galaxies belonging to these clusters have smooth appearances with 
no spiral structure, as such we classify all featureless disks as S0s,
regardless of bulge-to-disk
ratio\footnote{Yet, in the Hubble sequence S0 galaxies have more prominent
bulges relative to spirals \citep[e.g.][]{dressler80}.}.  We note that the
lack of spiral structure may be a resolution issue.  Our typical $V$-band
seeing is $1.2\arcsec$ FWHM corresponding to physical resolutions
of 910 pc (A85, A754) and 540 pc (A496); therefore, while we resolve the
cluster members, galaxies with weak spiral
arms may look smooth at the distances of these clusters.  We account
for this possible bias in our quantitative analysis of morphology in Paper~1.
Our purpose here is to provide basic visual morphology 
information for our analysis of different cluster populations based on
well-defined color criteria, and we present many examples in Figures
\ref{stamps85}-\ref{stamps754}.
We discuss the relative ratios of different morphological types as a function of
galaxy color in \S \ref{CMmorphs}.  In Paper~1,
we address differences in quantitative measures of morphology,
such as bulge-to-disk ratio and disk substructure, among different color-based
galaxy populations from these clusters.

In column (7) of Table~7 we give the fraction of basic morphological
types among cluster galaxies brighter than $0.1L^{\star}$.  These morphological
ratios are in agreement with the bulk of those from global cluster populations
studied by \citet{dressler80}.
The overall numbers of bright E, S0, and S$+$Irr galaxies in these clusters
are $163\pm33$, $311\pm87$, and $72\pm14$, respectively, which correspond
to the relative fractions of 0.30/0.57/0.13 (E/S0/S+Irr) morphological
types from the total bright cluster membership.

\subsection{Rest-frame $(U-V)$ Galaxy Colors}
\label{colors}
We require precise $(U-V)$ color measurements for every cluster galaxy.
With these we will construct C-M diagrams and determine empirical CMRs
for each cluster.  In this section we describe our cluster
galaxy color measurements including our choice
of aperture size, and the necessary photometric
corrections and their corresponding uncertainties.

The effects of the variable PSF over
each individual Mosaic
image, plus differing seeing conditions between $U$ and $V$ observations, 
combine to
adversely affect the image quality of each individual galaxy by different
amounts.  We account for varying image quality by degrading
all $U$ and $V$ galaxy imaging to a common PSF size for each cluster:
2.5 (A85), 2.0 (A496), and 2.5 (A754) arcseconds (FWHM).  
This correction insures
that we measure galaxy aperture magnitudes for colors in a consistent manner.
We use {\sc daophot} to construct a variable PSF model for each cluster
mosaic from hundreds of stars distributed evenly over the image.
We obtain the PSF size (FWHM) at each galaxy's location in our imaging from
the PSF model.  The image quality
characteristics of each combined cluster frame varying only a few percent
except near the outer image edges where focal plane
distortions start to dominate.
The typical PSF size for the images is roughly $1\farcs3$ FWHM, with a tail
of larger values.  We degrade each image to the largest PSF size 
using a Gaussian smoothing kernel.  While some information may be smeared out 
in the bulk of galaxies found away from mosaic edges, our procedure produces
uniform galaxy images for consistent aperture color measurements free of
systematic biases due to different PSF sizes.  We note that the total
integrated flux we use to derive absolute magnitudes
are derived from PSF-convolved fits to
surface brightness profiles, which require no image quality correction.

To calculate the $(U-V)$ color of each galaxy we measure aperture magnitudes
from the $U$ and $V$ images smoothed to a common PSF size.  
We select a fixed aperture size to directly compare our $(U-V)$ color data
with that of Coma from \citet[][hereafter BLE92]{bower92a,bower92b}.
For Coma ($cz = 6500$~\kms), BLE92 used an $11\arcsec$ aperture
diameter\footnote{This size corresponds to their $13\arcsec$ aperture
with a small correction applied to match the same physical size as the
$60\arcsec$ aperture they used for Virgo cluster galaxies 
\citep[see][]{bower92a}.}.  We scale the BLE92
aperture by the ratio of the line-of-sight
comoving distance to Coma and to each of our clusters to
obtain comparable fixed aperture diameters of $4.45\arcsec$ (A85),
$7.42\arcsec$ (A496), and $4.50\arcsec$ (A754).
Therefore, the fixed aperture we use
to measure member $(U-V)$ colors in each cluster samples the same physical
diameter of $3.34$~\hkpc in a $\Lambda$-CDM
($\Omega_{\rm M}=0.3$, $\Omega_{\rm \Lambda}=0.7$) cosmology.

We measure aperture fluxes using the {\sc apphot} package in IRAF.
This package is the standard for performing
aperture photometry on uncrowded digital images.
To ensure the sky is measured well outside
each galaxy we use an annulus with a width of 5 pixels and an inner radius
at least six times larger than the galaxy's half-light radius.
Our typical sky annulus contains $1000-5000$
pixels, with a minimum of $\sim300$ pixels for the smallest galaxy
images.  The local sky in our images is quite flat
given the steps we have taken to make the images globally flat (\S \ref{Redux}).
For each final cluster image we calculate the mean sky level in 48 separate
$30\times 30$ pixel boxes, free of celestial objects and bad pixels, and
find flat-field uncertainties of $0.8-1.1\%$ in $V$, 
and $0.9-1.6\%$ in the $U$-band.

We transform the aperture magnitudes to the \citet{landolt92} photometric
system by solving the
following color-correction equations iteratively:
$V_i = V + \beta_V (U_{i-1} - V_{i-1})$ and
$U_i = U + \beta_U (U_{i-1} - V_{i-1})$, where
$U$ and $V$ are the uncorrected aperture magnitudes from
equation~(\ref{mageq}), and the $\beta$ coefficients are color calibration
terms given in Table~4.  For the initial iteration we set
$(U_{i-1} - V_{i-1}) = (U-V)$.  We perform these calculations until
the magnitude difference between successive iterations is
$\delta m \leq 0.001$~mag.  The resultant average color corrections
are -0.04 (A85), -0.02 (A496), and -0.04 (A754) mag.

Using Galactic reddening values from SFD98 we apply a $(U-V)$
extinction correction of 
$A_{U-V}=2.119\times E(B-V)$ to each cluster galaxy $(U-V)$ color.
The mean corrections are 0.08 mag (A85), 0.27 mag (A496), and 0.14 mag (A754).
Finally, we apply $k$-corrections of -0.05 mag (A496) and -0.08 mag (A85,A754),
on average,
from \citet{poggianti97} to produce the final rest-frame $(U-V)$ colors.

We note that biases are associated with the choice of aperture size used to 
measure the colors of galaxies.  Fixed circular apertures do not take
into account the different physical sizes of galaxies, which may affect
the measured CMR because brighter galaxies have larger sizes 
\citep{scodeggio01} and color gradients are anticipated in most types of
galaxies \citep[see][and references therein]{peletier90,dejong96}.
Therefore, we test the effects that three different aperture sizes 
(i.e. fixed vs. multiples of the half-light radius $r_{\rm hl}$)
have on the measured CMR properties of each cluster in Appendix \ref{apertures}.

Our $(U-V)$ color measurements have formal random errors of
$\approx0.02-0.05$ mag.  The Poisson noise in the aperture magnitudes is
the dominant contribution, while additional sources include the color
correction, flat fielding, and dust ($10\%$ in $E(B-V)$) uncertainties.
All cluster members are bright enough that errors due to dark current
and read noise are not significant.  
We present galaxy colors from the three aperture choices for a small 
sample of cluster members
in Table~6; the full catalog is available electronically.
In addition, there are $(U-V)$ systematic error sources from
the overall photometric zero point (0.04 mag), the relative extinction zero
point (0.04 mag), and the $k$-correction (0.01 mag).

\begin{figure}[hp]
\center{\includegraphics[scale=.73, angle=0]{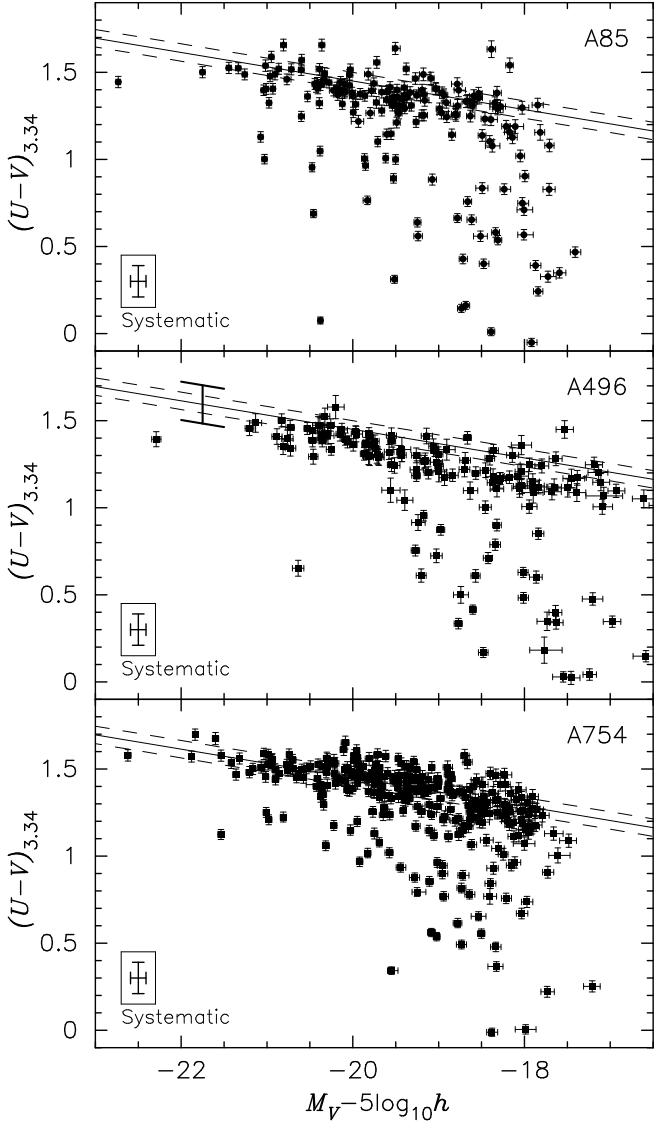}}
\caption
{Rest-frame $U,V$ C-M diagrams for the total sample of 637 spectroscopic members
in clusters A85, A496, and A754.  The
colors and total absolute magnitudes are fully corrected (see text for details).
The colors are from fixed aperture magnitudes with $3.34$~\hkpc physical 
diameters.  The systematic photometric
uncertainties are given at lower left, and the random errors for each galaxy
are shown with $\pm1\sigma$ errorbars.  For each cluster we plot
the fit to the CMR of E/S0s in Coma from BLE92 (thin solid line),
and we include the $\pm1\sigma$ intrinsic scatter of 0.050 mag (dashed lines).
The red-sequence galaxies in clusters A85 and A754 are well-fit by the
BLE92 result.  Cluster A496 is also consistent with the BLE92 fit when
the 0.11 mag systematic error (large bold error bar at $U-V=1.60$) in 
CMR zero point is taken into consideration.
\label{TheCMDs_fixap} }
\vspace{-0.2cm}
\end{figure}

\section{Results and Discussion}

The goals of this study are two-fold: (1) to establish the properties of 
the CMRs of local massive clusters and to test whether they
are universal; and (2) to search present-day
clusters for newer members.  For both aims we have opted for observing
cluster galaxy populations in $(U-V)$, $V$ C-M space.
Simple population synthesis models 
\citep[see e.g.][]{worthey94,vazdekis96,kodama97,bruzual03} show that
an $U$-band inclusive color provides the largest leverage for distinguishing
galaxies with relatively recent ($<2$~Gyr ago) episodes of SF
and, therefore, for estimating both the
range of stellar ages in red-sequence galaxies \citep{bower98}
and the membership age using blue color relative to the CMR
\citep{balogh00,kodama01a,bicker02}.
We note that the \citet{worthey94} age-metallicity degeneracy
($\Delta [Fe/H] \sim 2/3 \Delta \log{t}$) adds scatter to an absolute
correlation between blue galaxy colors and
young stellar ages; nevertheless, it appears that the CMR in clusters
is a metallicity-mass (i.e. metallicity-magnitude) relation predominately
\citep{kodama97,kauffmann98,vazdekis01}.  Furthermore, a tight relation between
metallicity and luminosity is found in galaxies with significant SF
\citep{skillman89,vanzee97,garnett02}.
Hence, at a given magnitude we do not expect vastly different metallicities
associated with blue colors, but rather different stellar ages.  Clearly,
quality spectra would validate our expectation between blue color and
galaxy age, yet such observations are beyond the scope of this particular
work.  In this section, we first fit each cluster's CMR to determine the
slope, scatter, and zero point.  Then we quantify the uniformity of the CMR
properties and identify the most recently accreted cluster members.

\subsection{Maximum Likelihood Fit}
\label{cmrLike}

In Figure~\ref{TheCMDs_fixap} we plot the rest-frame C-M diagrams for
the full spectroscopic memberships of clusters A85, A496, and A754.
These diagrams are based on colors from fixed aperture diameters
with the same physical dimension ($3.34$~\hkpc) as used by BLE92 for Coma.
The CMR is apparent in each cluster as the envelope of red galaxies
(the red sequence) comprising the bulk of the cluster membership.
In addition, we see that a fraction of the cluster members 
have colors $0.5-1.0$ mag bluer than the CMR.

We quantify the cluster CMR by modeling it
as a straight line in C-M space:
\begin{equation}
(U-V)_{\rm mod} = \frac{d(U-V)}{dM_V} (M_V - 5\log_{10}h + 20) + (U-V)_0 ,
\label{CMREqn}
\end{equation}
with slope $\eta_{\rm CMR}=d(U-V)/dM_V$, and zero point 
$(U-V)_0$ such that this is the
CMR color at a fixed absolute magnitude $M_V=-20+5\log_{10}h$.
Following the procedure of \citet{rix97}, we apply a maximum-likelihood
analysis to determine the best-fitting parameters
[$\eta_{\rm CMR},(U-V)_0,\sigma_{\rm CMR}$], and their associated
uncertainties, which describe the CMR with some intrinsic scatter
$\sigma_{\rm CMR}$ along the color direction.

We assume the model probability
distribution $P_{\rm mod}$ of the CMR is a Gaussian
of the form:
\begin{equation}
P_{\rm mod} (U-V) = \frac{1}{\sqrt{2\pi} \sigma_{\rm CMR}} \exp \{ - \frac{[(U-V) - (U-V)_{\rm mod}]^2}{2 \sigma_{\rm CMR}^2} \} .
\label{CMRModProb}
\end{equation}
Similarly,
the probability distribution of the ``true'' CMR, given a galaxy magnitude
$M_{V,i}\pm \sigma_{M,i}$ and color $(U-V)_i \pm \sigma_{U-V,i}$ observation,
is approximated by
\begin{equation}
\scriptstyle
P_{\rm obs} [M_V,(U-V)] = \frac{1}{2\pi \sigma_{M,i} \sigma_{U-V,i}} \exp \{ - \frac{(M_V - M_{V,i})^2}{2 \sigma_{M,i}^2} - \frac{[(U-V) - (U-V)_i]^2}{2 \sigma_{U-V,i}^2} \} .
\label{CMRTruProb}
\end{equation}
For the model parameters [$\eta_{\rm CMR},(U-V)_0,\sigma_{\rm CMR}$], 
the probability of making the galaxy photometric observation
$\{M_{V,i} \pm \sigma_{M,i}$ and $(U-V)_i \pm \sigma_{U-V,i}\}$
is given by the product of equations (\ref{CMRModProb}) and (\ref{CMRTruProb})
integrated over all
color and absolute magnitude space:
\begin{equation}
\scriptstyle
P_i [M_{V,i},\sigma_{M,i},(U-V)_i,\sigma_{U-V,i}] = \int_0^{\infty} \int_0^{\infty} P_{\rm mod}
\cdot P_{\rm obs} d(U-V) dM_V .
\end{equation}
Finally, the likelihood $\mathcal{L}$
of a CMR, given $N$ galaxy observations, is then
\begin{equation}
\mathcal{L} [\eta_{\rm CMR},(U-V)_0,\sigma_{\rm CMR}] = \sum^N_{i=1} \ln{(P_i)} .
\label{CMRLike}
\end{equation}

We find
the best-fit parameters [$\eta_{\rm CMR},(U-V)_0,\sigma_{\rm CMR}$] defining the
CMR and its intrinsic scatter by
maximizing the likelihood in equation (\ref{CMRLike}).  For a set of cluster
galaxy observations, we apply this maximum-likelihood technique in an
iterative fashion.  Initially we fit all galaxies in a given
sample.  The cluster red sequence is a well-defined correlation; therefore,
even with outliers included, our maximum-likelihood fitting finds the
roughly correct general CMR form (slope and zero point), but with a large
intrinsic scatter.  For the 
remaining iterations we use the previous $\sigma_{\rm CMR}$
to $\pm 3 \sigma$ clip the predominately blue outliers from consideration during
each fit.  In their C-M analysis, \citet{gladders98} found that linear fits
with $3 \sigma$ clipping provided the most stable results.  Similarly,
our procedure quickly converges to a stable best-fit CMR with a 
well-defined intrinsic scatter, requiring eight sigma-clipping iterations
on average.

We note that each galaxy's uncertainty in observed total magnitude
$\sigma_{M,i}$ does not play a crucial role in determining the most likely
CMR fit.  We test this by comparing CMR fits to cluster galaxy
photometry with and without $\sigma_{M,i}$ included.  We find the best-fit
results match within the model uncertainties.  In contrast, the size of the
observed color errors $\sigma_{U-V,i}$ directly relates to the best-fit
CMR such that smaller random errors in color produce larger values in the
model dispersion.  In other words, our method directly measures the
actual intrinsic scatter of the CMR.

The confidence interval for these
parameters are derived from the distribution of
\begin{equation}
\Delta \chi^2 = 2( \mathcal{L}_{\rm max} - \mathcal{L} ) .
\end{equation}
$\mathcal{L}_{\rm max}$ is the likelihood value for the best-fit CMR, and
$\mathcal{L}$ is the likelihood distribution about $\mathcal{L}_{\rm max}$
for each parameter.  We hold each parameter fixed at its best-fit value
and allow the remaining two parameters to vary, thus
each parameter's marginalized
$1\sigma$ confidence limits are given by the $\nu=1$ degree of
freedom condition $\Delta \chi^2 = 1$ \citep{press92}.

\subsection{Universality of the CMR}
\label{universality}
\subsubsection{CMR Properties of Four Nearby Clusters}

In the Introduction, we have made the case for the importance of 
determining the degree of uniformity among the CMRs of nearby clusters 
of galaxies.  Here
we apply the maximum-likelihood technique given above to quantify the
CMR properties of the three clusters in our sample.  To allow a
clean comparison of CMR properties, we restrict our cluster CMR samples
to match both each other and the BLE92 study of Coma by using the same
fixed color aperture, and comparable sample sizes and sampling radii.  
BLE92 fit the CMR of 48 E/S0s within a rough sampling radius
of 0.6 $h^{-1}$~Mpc centered on Coma's core.  For a direct comparison to this
external study, we test how both sample size and sampling radii affect the
measured CMR properties.

For each cluster galaxy we measure its color
in a fixed aperture diameter that corresponds to a metric size of
$3.34$~\hkpc matched to that used by BLE92 (see \S~\ref{colors}).
By using the BLE92 sampling radius of 0.6 $h^{-1}$~Mpc, we achieve
samples of 60 (A85), 91 (A496), and 95 (A754) red galaxies.  As a result
of the depth and high completeness of our cluster membership catalogs, these
samples are somewhat larger than the 48 E/S0s in BLE92.
We note that this physical sampling radius spans a range of 
virial radii fractions in our clusters; i.e. 0.37 to 0.50 $R_{200}$.
We also try reduced sampling radii that yield CMR sample sizes that are
comparable to that of BLE92.
In Table~8, we tabulate the results of our maximum-likelihood
best-fit CMR parameters for our clusters.
For each CMR fit we include information on the sampling radius and the total
number of galaxies inside, with and without outlier rejection.
BLE92 give the best-fit CMR parameters for Coma as follows:
$\sigma_{\rm CMR}=0.050$, $\eta_{\rm CMR}=-0.082\pm0.008$, and 
$(U-V)_0=1.45\pm0.11$.  We see that the CMRs of clusters
A85, A496, and A754 are very well-matched to that of Coma
when considering red galaxies with colors measured within the same
physical size apertures, and either similar sample sizes or sampling radii.

We find that the four clusters, spanning a fairly broad range in mass, have 
CMR properties confined to a tight range: intrinsic scatter [0.047,0.079],
slope [-0.094,-0.075], and zero point [1.35,1.45] (from rows 1, 5, 7, 12, 13,
and 19 of Table~8).
Clusters A496 and A754 have very tight $\sigma_{\rm CMR}$ values in good 
agreement with that of Coma, while A85 has a slightly broader CMR scatter
that is statistically consistent with the others.
Each cluster has a CMR slope that is identical within the measurement
uncertainties, and the average zero point color (at $M_V=-20+5\log_{10}h$)
of each cluster is well within the large errorbar of the Coma result.
The 0.07 mag spread among our zero point values can be
explained fully by our 0.09 mag systematic error in $(U-V)$ color;
therefore, these cluster CMRs are consistent with having homogeneous 
colors in agreement with \citet{andreon03}.

Furthermore, we find that the CMR
of each cluster is consistent with a simple, single slope model.
Contrary to \citet{metcalfe94}, who found a change in the ultraviolet CMR
slope at roughly 1 mag fainter than $L_B^{\star}$ using a large sample of
galaxies towards cluster Shapley 8, we observe no break
in any of the cluster CMRs down to at least $M_V=-18+5\log_{10}h$.
Our findings are based on individual, well-sampled CMRs from three Abell
clusters with large spectroscopic memberships, while the Metcalfe et al. 
result was based on a single cluster with relatively little redshift
information.

In Appendix \ref{apertures}, we show that some variance in
CMR properties occurs when considering $(U-V)$ colors measured in different  
apertures.  Regardless of sample selection, we find that using colors from
apertures containing 
the same fraction of galactic light does not remove the slope
of the CMR in clusters as claimed by \citet{scodeggio01}.  Nevertheless,
the half-light aperture choice for galaxy color
does expand the ranges of CMR scatter [0.053,0.112] and slope [-0.104,-0.054]
for the cluster populations within the same projected cluster-centric 
radius of $1/3R_{200}$ (see Table~8 rows 2-4, 8-10, and 15-18). 
We find differences of $2-3\sigma$
(measurement error) among the CMR properties for a single cluster when
comparing colors from apertures encompassing different fractions of the
total light.  As a result of red inward color gradients in early-type
galaxies \citep{franx89,peletier90}, larger color apertures produce
systematically bluer colors (i.e. include more
blue flux from the outer parts of galaxies) resulting in a systematic blueward
shift in CMR zero point, a somewhat larger intrinsic CMR scatter, and a slight
flattening of the CMR slope.  We illustrate these CMR variations in
Figure \ref{CMRcomp}, where we show the CMRs for each cluster using the
three different apertures for color measurements.  The maximum differences
in CMR parameters between the three clusters occurs when using half-light
apertures (middle row of Figure \ref{CMRcomp}).

\subsubsection{Constraints on Red-Sequence Stellar Populations}
We show that the $(U-V)$-based CMRs of local clusters are quite
uniform and robust to variations in color aperture selection, sampling
radius, and sample size.  As a result of the rest-frame $(U-V)$ color
straddling the 4000\AA\ break, we expect that the near-UV CMRs of the red
galaxies towards the centers of local clusters
are more sensitive to minor differences in SFHs than red CMRs, as illustrated
in the Introduction.
Yet, there are few studies in the literature of blue
or red rest-frame CMRs of local clusters.
We find that the range of intrinsic $(U-V)$ color scatter 
in our clusters is consistent
with previous results for low-redshift clusters 
\citep[e.g.][]{visvanathan77,terlevich01}.  At redder wavelengths,
\citet{pimbblet02} have studied 11 local X-ray luminous clusters in
rest-frame $(B-R)$ and found a somewhat broader scatter of 0.13 mag.
\citet{hogg04} reported a 0.05 mag scatter in $(g-r)$ for
bright SDSS galaxies residing in high density regions that are presumably
clusters.  Even though our $(U-V)$ analysis is in principle more sensitive 
to any recent star formation in galaxies on the CMR, we do not detect any 
scatter in excess of that from the redder CMR determinations.  
This result supports the picture that the core cluster population formed
long ago.

\citet{bower92b} concluded that the small intrinsic scatter in the $(U-V)$
CMRs of Coma and Virgo are due to a coeval and passively evolving population
of cluster early-type galaxies that formed before $z=2$.
Recently, this picture has been modified within the context of 
hierarchical structure formation such that the stellar populations of
present-day cluster early types are old ($z_{\rm f}> 1.0$) and
fading passively, while the galaxies in which they reside formed more
recently from merging systems \citep{bower98,vandokkum01a}.  Others have
reached similar conclusions that the stars in galaxies on the red sequence
formed long ago ($z>2$) based on the universality of the CMR slope in local 
clusters \citep{gladders98} and the homogeneity of the color of the red
sequence \citep{andreon03}.  \citet{bower98} showed that, under the 
assumption of a single SF event over a short timescale, the $(U-V)$
CMR scatter of Coma constrains the spread in ages of the bulk of the 
stellar population to $\Delta t \sim 4$ Gyr.

In light of the variety of reasons that contribute to
small variations in CMR scatter, our results are consistent with the
interpretation that a small intrinsic CMR scatter
corresponds to a small spread in stellar population age.  Following
the analysis of Bower et al. under the simple assumption of a single
burst model, the range we find in CMR scatter based
on fixed metric apertures (i.e. [0.047,0.079]) is consistent with a
spread in stellar ages of $1.4< \Delta t < 4.1$ Gyr over a range
of IMFs and metallicities (solar $\pm60\%$) 
using two population synthesis codes:
\citet{worthey94} and \citet{vazdekis96}.  This result is in accord with
the Bower et al. finding.  In a flat universe with
$H_0=71$~km~s$^{-1}$~Mpc$^{-1}$ and $\Omega_{\rm M}=0.27$, 
the maximum age spread of 4.1 Gyr
corresponds to a minimum formation redshift of $z_{\rm f}=1.6$.
If we consider the expanded intrinsic scatter range [0.047,0.112]
when using galactic half-light color apertures, we find that the maximum
stellar age spread increases to $\sim5.2$ Gyr ($z_{\rm f}>1.2$).
\citet{bower98} also considered continuous SF models and they found
that the tight CMR scatter of the Coma Cluster still places strong
constraints on the formation times of the bulk of stars, yet the last
epoch of SF is weakly constrained to no more than a few Gyr in the past.
The larger maximum values of CMR scatter that we report are consistent
with a small amount of SF continuing to the present day, and the
bulk of the stellar populations having old ($z_{\rm f}>1$) ages.

Another way to test the picture in which most of the stars in the 
red-sequence galaxies form early is to compare the scatter of the CMR
in low and high redshift clusters.  Rest-frame $(U-V)$ color scatters of
$\leq 0.09$ and $\sim0.07$ have been reported for over 20 clusters at $z>0.5$
\citep{ellis97,stanford98}.  The CMR scatters we have observed among our three
local Abell clusters are consistent with these high redshift values.
Therefore, we find no evidence for evolution of the $(U-V)$ CMR scatter
in clusters since $z\sim1$, a result consistent with the constraints
imposed above by the observed variation in the scatter for our nearby 
clusters.

\subsection{Identifying the Most Recently Accreted Cluster Galaxies}
\label{defPops}
 
Finding the most recent arrivals from the field is key
to testing the hierarchical picture of cluster formation and for
follow-up studies of what factors might influence the evolution of   
galaxies in cluster environments.  Here we use a combination of 
spatial, kinematic, and morphological data to extract this infalling 
population.
With our precise and well-defined CMRs, we first divide each cluster galaxy
population into three subsamples based on individual galaxy color difference
$\Delta (U-V)$ with respect to its default CMR as follows:
\begin{enumerate}
\item Red sequence galaxies (RSGs) with $\Delta (U-V) \geq -2 \sigma_{\rm CMR}$.
\item Intermediately blue galaxies (IBGs) with
$-2 \sigma_{\rm CMR} > \Delta (U-V) > -0.425$~mag.
\item Very blue galaxies (VBGs) with $\Delta (U-V) \leq -0.425$~mag.
\end{enumerate}

For the purpose of producing well-defined samples of cluster members
separated in C-M space, we adopt the use of half-light
aperture colors for defining the default CMR in each cluster because this
selection accounts objectively for the different sizes of galaxies.
Furthermore, 
since the CMR parameters of each cluster appear essentially independent 
of sample size and sampling radius, we will select red galaxies within
identical physical radii to define the default CMRs of the clusters.
We choose $1/3R_{200}$ for the following
reasons: (1) it is representative of the cluster ``core'' where the
red galaxies are most centrally concentrated and the number of blue
outliers are minimal; (2) it is midway between the sampling radii we
used above in our comparisons with the Coma CMR results;
and (3) it is a well-defined fraction of the projected
virial radius.  We adopt a $\sim10\%$ uncertainty in $R_{200}$ based on
typical errors in cluster redshifts ($\sim5\%$) and velocity dispersions 
($\sim5\%$).  In Table~8, we see that the best-fit CMR 
parameters for each cluster are statistically equivalent when using
sampling radii within $\pm10\%$ of $1/3R_{200}$.
Moreover, these parameters are in good
agreement with our quoted range above from the comparison with Coma.

Blueward deviations relative to the CMR are thought to be
driven mainly by differences in the relative ages of the constituent
stellar populations \citep[e.g.][]{vandokkum98,terlevich99}, 
which are in turn linked to time since accretion
\citep{balogh00,kodama01a,bicker02}.  Therefore,
these color-based types should 
provide a coarse, yet well-defined, sequence of time in
cluster residence; e.g., RSGs have long been cluster members, 
while VBGs with the integrated colors of star-forming spirals are likely
still in the process of infall.  We have used $(U-V)$ colors owing to their
greater sensitivity to relatively recent ($<2$~Gyr ago) star formation, which
gives us the best photometric leverage for separating galaxies comprised of
differing stellar ages.
We note that we have made no attempt to
use our visual classifications to separate the cluster samples into
different bins.  Our sequence of RSG, IBG, and VBG is based purely on
position in C-M space.

\begin{figure}[hp]
\center{\includegraphics[scale=.73, angle=0]{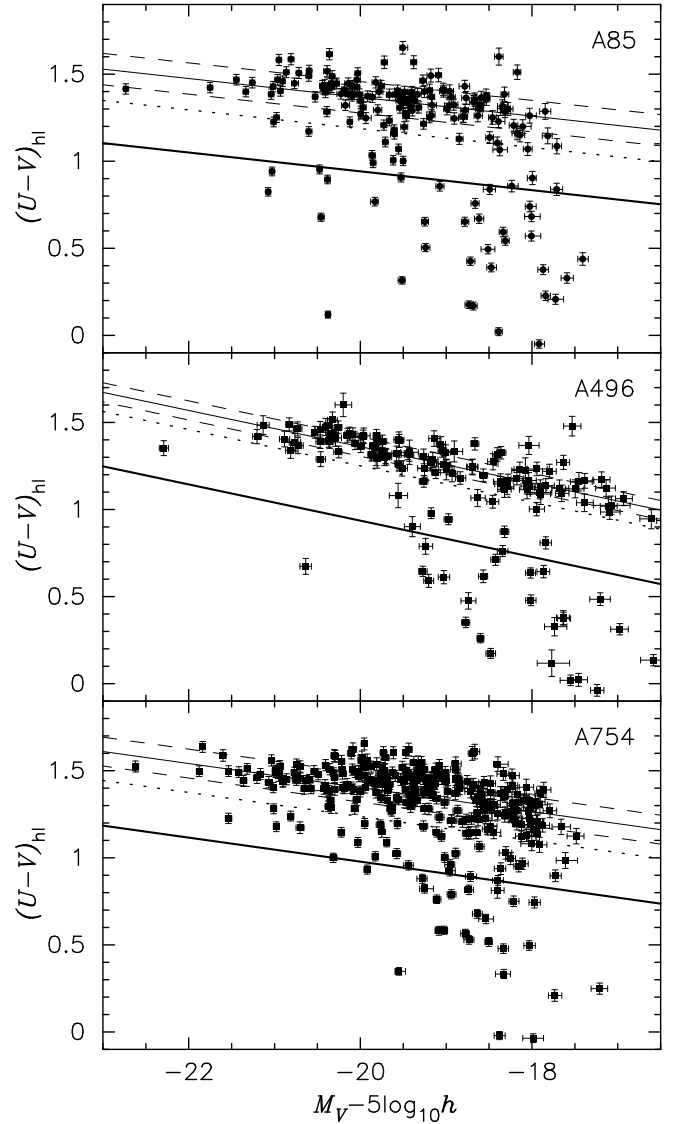}}
\caption
{Rest-frame $U,V$ C-M diagrams based on half-light aperture colors.  We
plot the full set of 637 spectroscopic members with the C-M population
boundaries shown.  For each cluster we give
the default CMR (thin solid line) and its $\pm1\sigma$ intrinsic 
scatter (dashed lines) based on our maximum-likelihood fit to the subset of
members within a projected cluster-centric radius of $1/3R_{200}$.
We use the $-2\sigma_{\rm CMR}$ limit (dotted line) to define the
separation between the blue and red galaxy populations.  The blue galaxies
are further divided into intermediately blue (IBG) and very blue (VBG) members
using a $\Delta(U-V)=-0.425$~mag criterion (bold solid line) that is 
equivalent to the $\Delta(B-V)=-0.2$~mag cut of \protect\citet{bo84}.
\label{TheCMDs} }
\vspace{-0.2cm}
\end{figure}

We draw the boundaries between these populations in the C-M diagrams presented in
Figure \ref{TheCMDs}.
The majority of cluster members belong to the
red sequence population found above the dotted sloped line representing
$\Delta (U-V) = -2 \sigma_{\rm CMR}$.  Below this line the blue members are further
divided into IBG and VBG types by the bold sloped line at $\Delta (U-V)=-0.425$,
which is similar to the \citet{bo84} criterion of $\Delta (B-V)=-0.2$ for
finding galaxies with colors significantly different from E/S0s 
(i.e. spiral-like).  We base our $(U-V)$ cut on the CMR behavior of
early-type galaxies illustrated convincingly by 
\citet{schweizer92} in a detailed study of $\sim500$ E/S0s
drawn from the Third Reference Catalog \citep{rc3}.  By inspection 
of their CMRs \citep[][Fig. 1]{schweizer92}, we find that blue 
cutoffs of $\Delta(U-V)<-0.425$~mag and $\Delta(B-V)<-0.2$~mag likewise
exclude nearly all E/S0
galaxies over a luminosity range ($-23.5<M_B<-18$)
very similar to our data.  This value is somewhat smaller than the
$\Delta(U-V)=-0.54$~mag Butcher-Oemler
criterion adopted by \citet{kodama01a}.  
We define the moderately blue IBG population with the aim of 
locating galaxies belonging to clusters for an intermediate time between the
most recent arrival VBGs and the long resident red galaxies.  We choose
$-2 \sigma_{\rm CMR}$, rather than $-3 \sigma_{\rm CMR}$, to separate
RSGs and IBGs for two reasons: (1) empirically this cut is a more natural
match to the blue envelope of the red sequences; and (2) this cut is closer to 
midway between the default CMR and the \citet{bo84} criterion for two of our clusters
and, thus, produces reasonable IBG sample sizes for analysis.

We give the C-M based population breakdown for each cluster in Table~9,
and we find $18-23\%$ of all cluster galaxies more luminous than $0.1L^{\star}$
belong to the IBG or VBG populations.  This finding is not in conflict with
the low blue (Butcher-Oemler) fractions found in local clusters
\citep[e.g.][]{bo84,rakos95,margoniner01} because those studies include
a luminosity cut that, if used here, would reduce our fractions of blue
galaxies to a few percent.  We are interested in all cluster galaxies
down to our completeness limits.

In the hierarchical picture of cluster formation, clusters grow through
accretion of galaxies from low density regions
\citep[][and references therein]{white78,white91,west95}.
If our three color-based types provide a coarse sequence of time since cluster
infall, then the bluer cluster members should show additional evidence
suggesting more recent infall, such as spatial and kinematic
differences compared to the red members \citep{diaferio01}.  
Therefore, we
examine basic properties, such as spatial distributions, kinematics
and morphologies, of each color-based population of cluster galaxies.

\subsubsection{Spatial Distribution Comparisons}
\label{spatial}
\begin{figure*}[hp]
\center{\includegraphics[scale=0.75, angle=0]{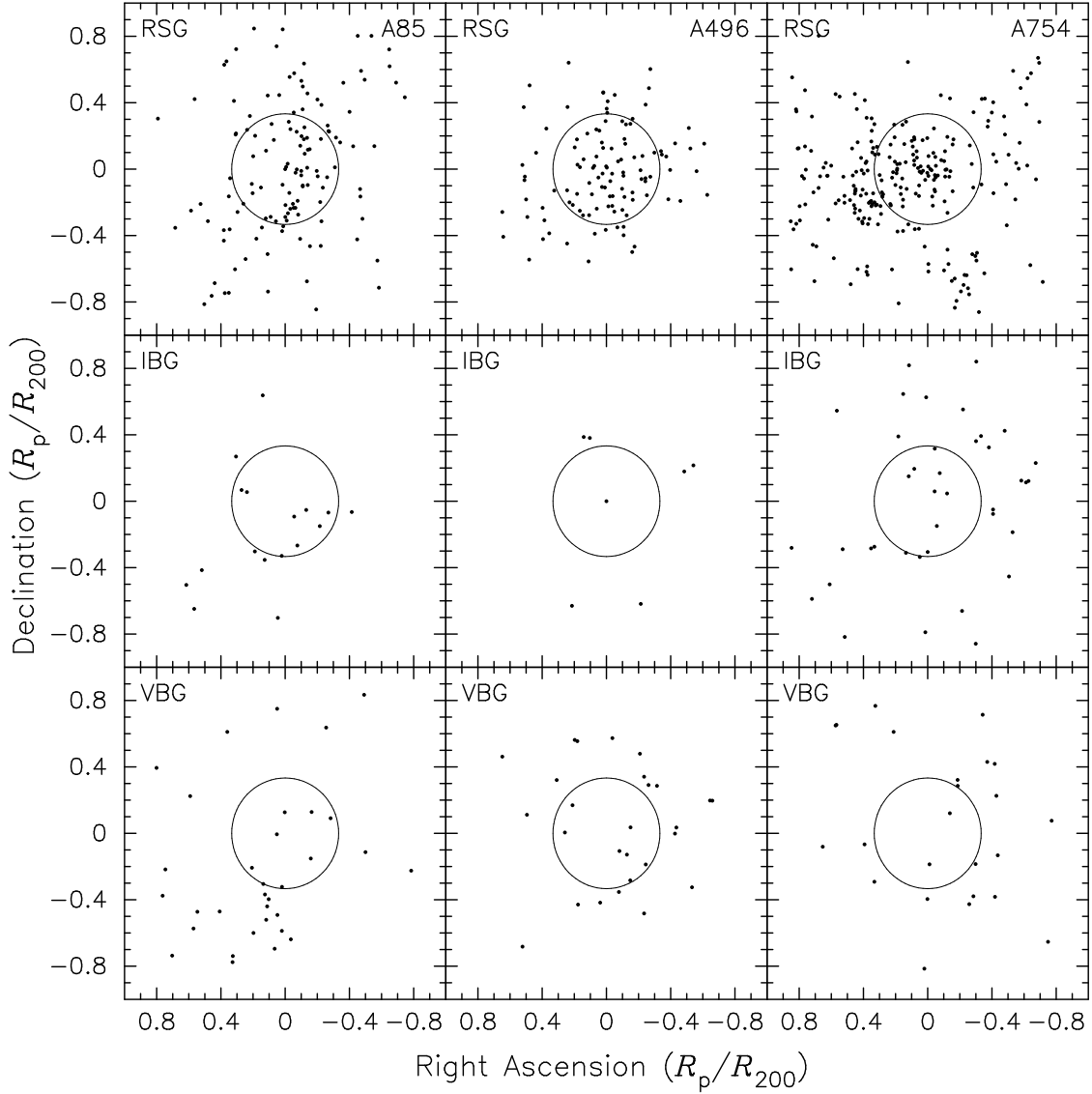}}
\caption{
Spatial distributions for color-selected cluster galaxy populations:
RSG (top panels), IBG (middle panels), and VBG (bottom panels).
In each panel we plot the projected cluster-centric distances \rp
relative to the brightest cluster galaxy for each spectroscopic member.
The \rp distances are given as a function of the estimated virial
radius $R_{200}$ of each cluster.  The spatial coverage of A496
is smaller than that of A85 and A754
because of its closer distance and our one square-degree imaging.
Each panel shows an $R_{\rm proj}=1/3R_{200}$ circle to mark the
region that defines the default CMR sample.  All panels
are oriented with north up and east to the left.
\label{space9} }
\vspace{-0.2cm}
\end{figure*}

In Figure \ref{space9} we plot the separate projected spatial
distributions of the three color-selected populations for each cluster.
The projected cluster-centric distance \rp to each member is equal to
the product of its angular separation from the cluster center (defined
by the brightest cluster galaxy),
and the line-of-sight comoving distance $D_{\rm C}(z)$ in a flat cosmology
\citep[see][]{hogg00}, where $z$ is the cluster redshift.  We show
\rp values in terms of $R_{200}$.  We note that our one-square-degree
imaging produces a somewhat truncated spatial coverage of A496 compared with
the other, more distant clusters.  From the field-of-view diameter given
in Table~2, we calculate the percent of projected virial radius
$R_{200}$ that our imaging covers for each cluster: 81\% (A85), 65\% (A496),
and 84\% (A754).  Naturally, we do observe some regions of each cluster
at slightly larger \rp corresponding to the corners of our imaging.
We see that there are concentrations
of RSGs (top panels) towards the cluster centers in contrast with the
blue populations (IBGs=middle, VBGs=bottom), which reside preferentially
outside the projected center.  Moreover,
the spatial distributions of red members are clumped or clustered, while
the blue galaxies appear more spread out.  The qualitative
segregation of cluster galaxies divided by color has been reported
\citep[see e.g.][]{ramirez00}.

\begin{figure*}[hp]
\center{\includegraphics[scale=0.85, angle=0]{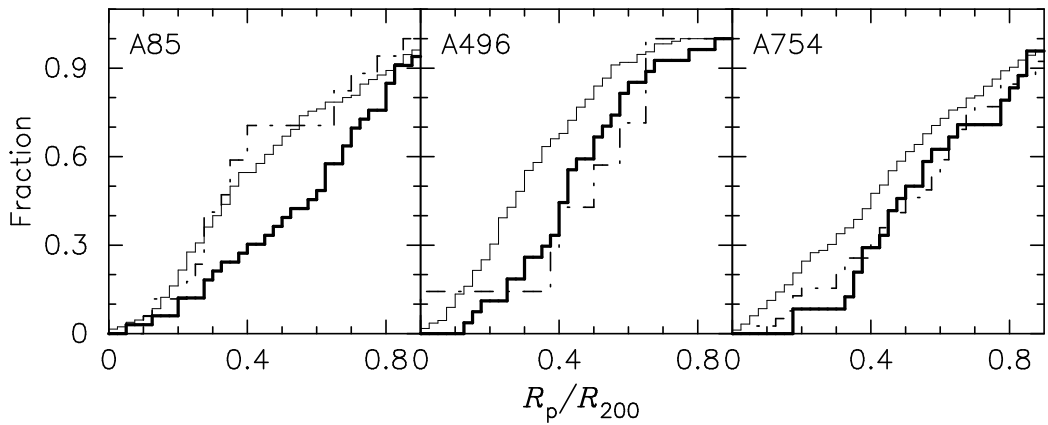}}
\caption{
Cumulative fractions of each color-selected population as a function of
cluster-centric projected radius as follows: RSG (grey line), IBG
(dot-dashed line), and VBG (bold solid line).
\label{cumul} }
\vspace{-0.2cm}
\end{figure*}

The spatial segregation between red and blue cluster members is most
apparent in all three clusters when comparing RSGs and VBGs.  Very few
VBGs lie within $1/3R_{200}$, and those that do are likely seen in
projection \citep{diaferio01}.  Among the IBG spatial distributions,
only that of A754 obviously avoids the inner cluster, and it resembles
its VBG population.  A496 has too few IBGs, and those in A85 are more
centrally concentrated than the very blue members.
We illustrate further the differences
in relative locations by showing
cumulative spatial distributions for the color-selected samples in
Figure \ref{cumul}.  For example, less than 30\% of the VBG population
in each of our clusters is found inside of $R_{\rm p}=0.4R_{200}$.
In contrast, 50\% or more of the red membership is found inside the same
radius.  We point out that in A754 the fractions of red and blue members
inside $0.4R_{200}$ are more comparable than in A85 and A496 due to the large
number of RSGs in the post-merging SE clump \citep{zabludoff95}.
For the overall cluster sample we find the blue galaxies have $R_{200}$
distributions that are quite different ($>99.9\%$ VBG and 97.8\% IBG, with
a K-S test) from the RSG members.
Finally, we note that the change in relative numbers of blue
to red members per radial interval is responsible for the
cluster-centric distance dependence of the Butcher-Oemler effect observed
by \citet{ellingson01}.

\subsubsection{Kinematic Comparisons}

\begin{figure}[hp]
\center{\includegraphics[scale=0.85, angle=0]{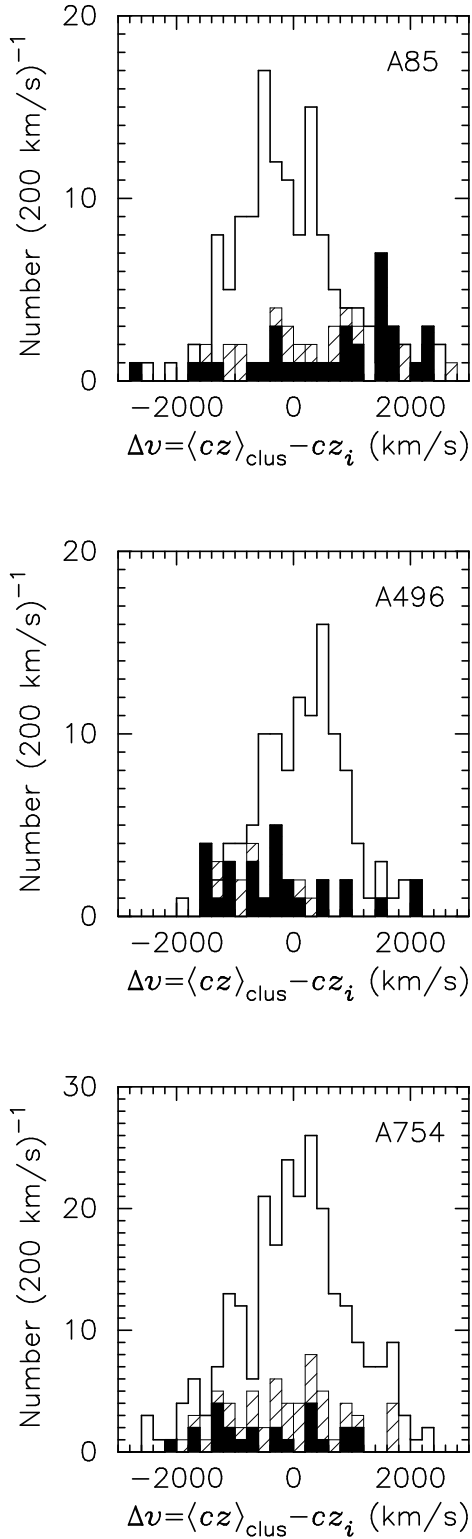}}
\caption{
Relative velocity histograms for color-selected cluster galaxy populations:
RSG (solid outline), IBG (hatched histogram), and VBG (filled
histogram).  We bin the IBG histograms on top of the VBG bins so that
both populations are unobscured.  The $\Delta v$ bin size is 200 \kms.
Each galaxy we plot is a cluster member defined to be within 
$3\sigma_{\rm clus}$ of $\left<cz\right>_{\rm clus}$ (see text).
Statistical test results from comparisons between the blue and RSG
populations are tabulated in Table~10.
\label{kinematics} }
\vspace{-0.2cm}
\end{figure}

In Figure \ref{kinematics} we show the relative velocity histograms
of cluster members divided into our three color-selected populations.
Relative velocities $\Delta v$ are given by the difference between the cluster
mean recessional velocity $\left<cz\right>_{\rm clus}$
(see Table~7) and the individual galaxy velocity
$cz_i$ from the redshift survey of \citet{christlein03}.  Thus, $\Delta v$
represents each galaxy's velocity relative to the cluster rest frame.  
The RSG populations of each cluster have velocity distributions that
are approximately Gaussian and centered near $\Delta v=0$.  These 
characteristics suggest the red members are gravitationally bound in
at least semi-equilibrium, in agreement with their centrally concentrated
spatial distributions.  In contrast, using combined samples from the three
clusters, a K-S test shows that the velocity distributions of the VBGs
and the red galaxies differ at the 95\% level.

We see in Figure \ref{kinematics} that 
the VBG samples have $\Delta v$ distributions
that are roughly flat and, in most cases, shifted relative to $\Delta v=0$.
The kinematics of the IBGs appear less distinct from the RSGs.
For each cluster we use several statistics to test how 
different the kinematic distributions of the blue galaxy populations 
are compared to the RSGs.  We tabulate the mean
velocity and the velocity dispersion for each color-selected galaxy
population in Table~10.  First we use a
K-S test to quantify the significance of differences in overall
velocity distributions, and we find that the VBG populations of clusters
A85 and A496 are significantly different than the red-selected galaxies.
We repeat these tests for the IBG populations in the two clusters (A85 and
A754) with reasonable sample sizes, and we find that the intermediate blue
members do not have significantly different kinematics compared with
their RSGs.  Next, to determine whether the kinematics of the VBGs exhibit
evidence for substructure or asymmetric infall, we test for
differences in the means and variances of the velocities
\citep[see e.g.][]{zabludoff93}.  We start by using an F-test to establish
whether the variances of the VBG and RSG velocity distributions 
in a given cluster are
significantly different.  Depending on the outcome of this test, we then
use either the Student's T-test or the ``unequal variance'' T-test
to test the significance of any difference in velocity means
\citep[see][]{press92}.
We give the results of these statistical tests in Table~10,
which show that there are significant differences in
both the velocity means and dispersions when comparing VBGs and RSGs in
two of the three clusters (A85 and A496).  For these two clusters, the
VBGs have dispersions that are $\sim \sqrt{2}$ larger than for the red galaxies,
indicative of an infalling rather than relaxed (i.e. virialized)
population \citep{adami98}.  Furthermore, the significant mean
offsets of the VBG relative to RSG velocities correspond to substructure
or asymmetric infall \citep{zabludoff93}.  We note that
the lack of velocity difference in cluster A754 may be the
result of the recent collision between two subclusters within the past
$\sim1$~Gyr \citep{zabludoff95}.  Evidence of this merging event is
apparent in the bimodal nature of this cluster's RSG spatial
distribution (see Figure \ref{space9}).

The IBG relative velocities are consistent with those of the RSGs, which
provides a qualitative confirmation of longer time in cluster residence
as suggested by their intermediate colors.
No color information was used by \citet{christlein03} to define the mean
velocity and dispersion of each cluster,
so it is not surprising that the RSGs are close to $\Delta v=0$ as
they make up the bulk of cluster members.
Any minor shifts in RSG velocity zero points are due to larger offsets
in the blue members, which are characteristic of infalling populations
along the line-of-sight.  Taken together, the spatial and kinematic 
properties of the blue galaxies provide compelling evidence 
supporting the idea that their color reflects recent or ongoing
accretion onto these clusters following the model predictions of
\citet{diaferio01}.

\subsubsection{Morphological Comparisons}
\label{CMmorphs}
Given the spatial and kinematic confirmation that blue cluster members
are relatively late arrivals, an examination of the
morphological content of the different color-based populations is warranted.
We describe our visual classification of cluster members in \S \ref{mmakeup},
and we give the breakdown of E/S0/S$+$Irr types as a function of galaxy color,
split between $L\ge0.1L^{\star}$ and $L<0.1L^{\star}$, in Table~9.
Qualitative morphology is highly subjective, with dependencies on resolution
and surface brightness.  Even with a rudimentary classification sequence of \\
E/S0/S$+$Irr, a pair of experts will disagree between adjacent types in some
fraction of a given sample.  Therefore, we adopt a conservative uncertainty
of 20\% in our visual classifications such that one-fifth of E and S$+$Irr types
could be misclassified as S0, and summing quadrature gives a 28\% chance for
S0s to be either other class.  We stress that our by-eye classifications,
which are based on inspection of single passband images, are
unbiased by color information and were performed independent of 
the subsequent divisions into color-based populations.

\begin{figure*}[hp]
\center{
\includegraphics[scale=0.85, angle=0]{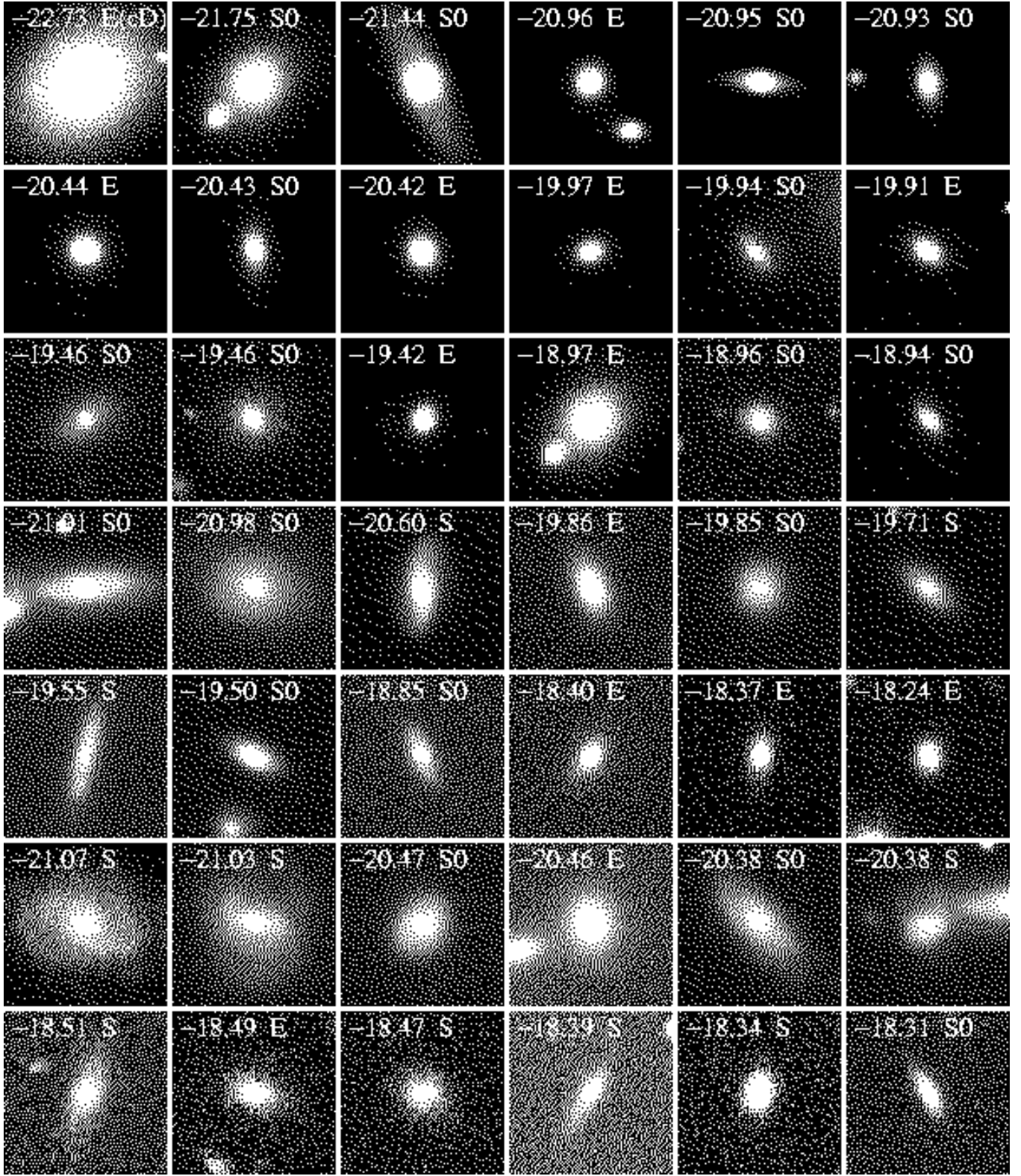}}
\caption
{Postage stamp $V$-band images of example members belonging to cluster A85.  We
show RSGs in the top three rows, with the 18 examples divided into six 
representative cuts in luminosity starting with the three most luminous, and
then decreasing in steps of 0.5 mag from the third brightest cluster RSG.
In the bottom four rows we give IBG (rows 4 and 5) and VBG (rows 6 and 7)
examples.  For the IBGs and VBGs we plot the six brightest in rows 4 and 6,
respectively.  Similarly, we plot the six faintest (with a minimum 
luminosity limit of $M_V=-18.1+5\log_{10}h$) in rows 5 and 7, respectively.
The total $M_V-5\log_{10}h$ magnitude and visual classification is given for 
each galaxy.  Each panel is 20~\hkpc on a side.
\label{stamps85}
}
\vspace{-0.2cm}
\end{figure*}

\begin{figure*}[hp]
\center{
\includegraphics[scale=0.85, angle=0]{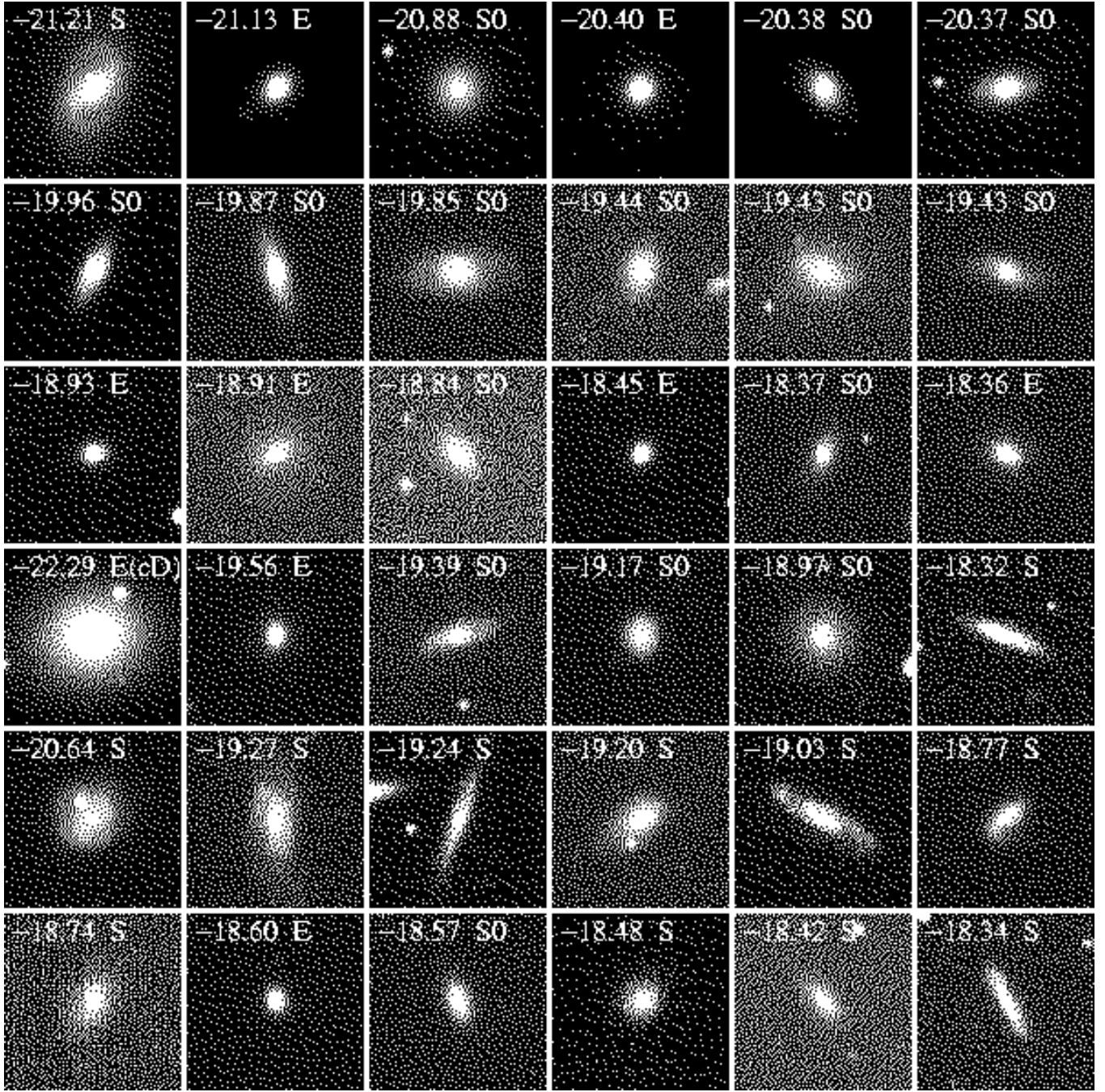}}
\caption
{Postage stamp $V$-band images of example members belonging to cluster A496.  The
representative RSG examples are plotted in rows 1-3 as in Figure \ref{stamps85}.
We show the six brightest IBGs (row 4) and the twelve brightest VBGs 
(rows 5 and 6).  Each panel is 20~\hkpc on a side.
\label{stamps496}
}
\vspace{-0.2cm}
\end{figure*}

\begin{figure*}[hp]
\center{
\includegraphics[scale=0.85, angle=0]{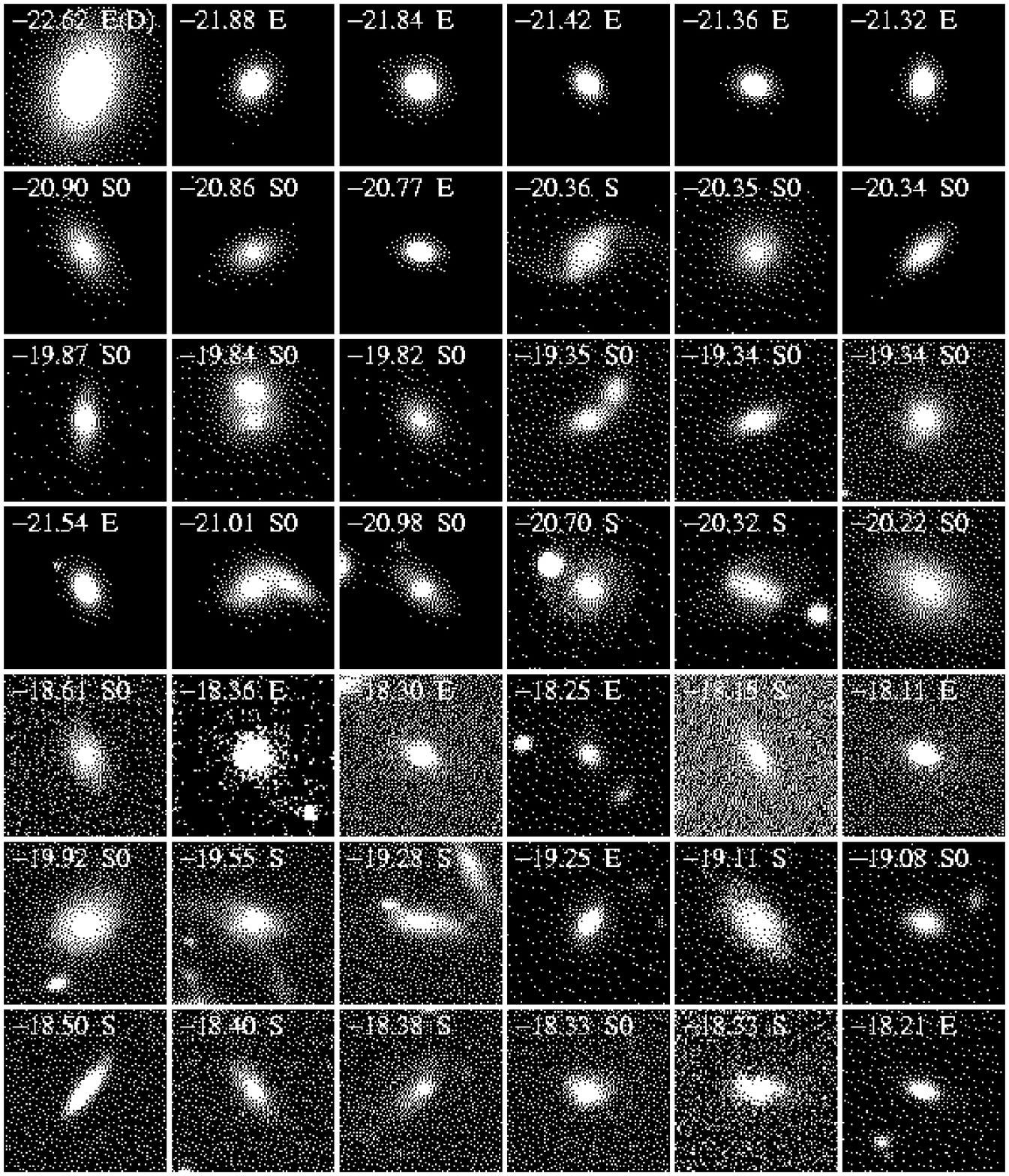}}
\caption
{We show example members belonging to cluster A754 following the same format
as in Figure \ref{stamps85}.
\label{stamps754}
}
\vspace{-0.2cm}
\end{figure*}

For the subsample of 546 cluster galaxies brighter than $0.1L^{\star}$, we find 
the relative numbers of three morphological types \\ (E/S0/S$+$Irr) for
each color-based population are
$142\pm28/270\pm76/24\pm5$ (RSG), $13\pm3/30\pm8/12\pm2$ (IBG), and
$8\pm2/11\pm3/36\pm7$ (VBG).  Bright RSGs are $94\pm17\%$ early-type (E/S0)
by number density; therefore, red early types make up three-quarters of the
galaxies inhabiting these low redshift clusters.  In the top three rows
of Figures \ref{stamps85}--\ref{stamps754} we show examples of RSGs divided
into six representative cuts in luminosity.  We observe a handful of red
spirals in each cluster -- examples include the brightest RSG of A496 
(Fig. \ref{stamps496})
and the $\sim L^{\star}$ barred spiral in A754 (Fig. \ref{stamps754}, -20.36 S).
These rare systems are likely examples of so-called ``passive spirals'' 
so far found
in or near cluster environments, which have spiral morphologies, red colors
and no SF activity \citep{couch98,dressler99,poggianti99,goto03}.

Luminous VBGs with the integrated colors of star-forming spirals are $65\pm13\%$
late-type (spiral or irregular) by number, and we classify another $20\pm5\%$
as S0s.  We give examples of VBGs in the last two rows of
Figures \ref{stamps85}--\ref{stamps754}.  
Most of the VBGs in our cluster sample are obviously disk
galaxies with small concentrated nuclei and apparent disks, yet many of those
classified as late-type have weak spiral structure, their features appear
smooth in contrast with archetype ``grand design'' galaxies.  Our admittedly
subjective visual classifications are based more on characteristics such as disk
lopsidedness or asymmetry, than strong spiral features.  
A detailed quantification of the disk substructure measured in these cluster
galaxies, including an accounting of selection effects, is a main aspect of
the analysis we present in Paper~1.  We note that \hst\ and deep ground-based
observations have subsequently shown that VBGs at redshifts $0.2<z<0.4$ are
mostly normal late-type spirals and irregular, possibly merging, systems
\citep{lavery94,couch94,dressler94,oemler97}.  Therefore,
the agreement between the morphological makeup of blue cluster galaxies at
$z\sim0.4$ and locally, combined with the decreasing accretion rate implied
by the Butcher-Oemler effect, are convincing evidence supporting the buildup
of clusters through the infall of field galaxies as expected in an
hierarchical framework.

There are a handful of ``ellipticals'' in the VBG sample 
(e.g. $M_V=-20.46+5\log_{10}h$ in A85, $M_V=-18.60+5\log_{10}h$ in A496, 
see Figures \ref{stamps85} and \ref{stamps496}).
It is not clear from our images whether these are true ellipticals or
abnormally luminous bulges of disk galaxies.  These morphologies are not
necessarily inconsistent with the latest arrival scenario, as there
are examples of blue ellipticals and blue bulges in the field
\citep[e.g. from the concentration of SF following a 
galaxy-galaxy interaction;][]{yang04}.

Finally, the moderately blue cluster population has a predominantly early-type
content with $78\pm15\%$ E/S0 and only $22\pm4\%$ S$+$Irr types.  IBG examples
are displayed in rows 4 and
5 of Figures \ref{stamps85} (A85) and \ref{stamps754} (A754); cluster A496 has
only six IBGs more luminous than $0.1L^{\star}$ and all are shown in the fourth 
row of Figure \ref{stamps496}.  We note the giant cD galaxy
at the center of A496 has moderately blue colors placing it in our IBG
population.  By numbers, the IBGs are over-abundant in E and S0
types, with an emphasis on S0s.  Nevertheless, we note that many IBG S0s
in these clusters have small concentrated bulges and
large smooth disks, which appear more like later-type disk galaxies with an
absence of the obvious spiral features found in grand-design, or even flocculent,
spirals.  Furthermore, we find similar S0s among the VBGs as well.  Some
examples of these possibly ``late-type S0s'' can be found in
Figure \ref{stamps85} (IBGs -21.01 and -20.98, and VBG -20.38) and
Figure \ref{stamps754} (IBG -20.22 and VBG -19.92).
We leave the detailed measurement of quantitative morphology to
Paper~1, where we account for possible resolution bias by artificially 
degrading a control sample of field galaxies for comparison,
and note here simply that the morphological makeup of the IBGs
may be intermediate between that of the red and very blue membership.

\subsubsection{Summary}
In this section we have shown that $\sim20\%$ of the
$\ge0.1L^{\star}$ galaxy population of local Abell clusters, 
located within $65-85\%$ of the projected virial radius,
have blue or moderately blue colors relative to the CMR defined
by the red members.  In general,
the cluster VBGs are spatially, kinematically, and morphologically
distinct from the more bound and centrally-concentrated red members with
smooth early-type morphologies.  In all three clusters, we find the VBGs
are more typical of the morphologies of star-forming field galaxies.
Moreover, the VBGs tend to lie at projected radii further
from the core, with typical velocities displaced from the cluster mean
velocity.  For two of the three clusters, the VBGs have velocity
dispersions that are $\sim \sqrt{2}$ times the dispersion of the red
members, which is expected for an infalling, as opposed to virialized, 
population.  These results imply that the bluest galaxies have entered
from the field in agreement with the model predictions of \citet{diaferio01},
and have not yet mixed with the redder (older) cluster galaxies in the core.
Our findings are in agreement with other studies that detect
a population of nonvirialized, late-type galaxies in clusters
\citep{zabludoff93,adami98,adami00}.
The presence of late infall members in low redshift clusters is 
expected in the hierarchical picture of cluster formation.
Identifying these galaxies is invaluable for follow-up
studies of what factors might influence the evolution of galaxies 
in cluster environments.  In Paper 1 we concentrate our
analysis on quantitative measures of morphology and structure 
for these late cluster arrivals to further understand their role in
the evolution of galaxies in cluster environments.

The spatial, kinematic, and morphological data suggest that an IBG
could represent a ``transition'' object between a more recent arrival with
young stars still in abundance (e.g. VBG)
and the RSG type now containing only an old stellar population.
IBGs make up $5-12\%$ of the galaxies in our cluster
sample.  Such a population provides evidence for the whereabouts of the blue
galaxies once prevalent in rich clusters.  The total fraction of
present-day cluster IBG$+$VBG populations is representative
of cluster VBG fractions at moderate redshift (e.g. $\sim20\%$).

\section{Conclusions}

We present a large data base of precise $U,V$ photometry for 637 present-day
cluster galaxies, most more luminous than $M^{\star}_V+3$.  The sample is
drawn from square-degree imaging of three nearby ($z<0.06$) Abell clusters:
A85, A496, and A754.  All
galaxies have cluster membership confirmed using spectroscopic redshifts from
\citet{christlein03}, which removes the need for uncertain field corrections.

To determine the degree to which cluster $(U-V)$ CMRs are uniform in the
local universe and thereby constrain galaxy evolution models,
we employ a maximum-likelihood technique to analyze the CMR properties of
the red galaxies within a projected radius of $1/3R_{200}$ for each 
cluster in our sample. 
Using a variety of red galaxy sample selections and $(U-V)$ color measurements,
we find that CMRs (defined by slope, intrinsic scatter, and zero point
at fixed absolute magnitude $M_V=-20+5\log_{10}h$)
remain quite uniform, with maximum variations in these properties as follows:
0.047 to 0.112 mag (intrinsic scatter); -0.104 to -0.043 (slope); and 
1.34 to 1.45 mag (zero point).  We note that the range in zero point color
can be explained by systematic photometric error, thus, our results are
consistent with homogeneous CMR colors in agreement with \citet{andreon03}.
The range of CMR scatter allows a maximum spread in the
formation times of the stellar populations of 5.2 Gyr, with the bulk
of the stars forming before $z=1.2$, following the analysis of
\citet{bower98}.

If we limit our analysis to the core red
galaxy populations and adopt the same color aperture as used for the
Coma Cluster by BLE92, we find that local clusters spanning a range of
cluster masses and richnesses have universal CMRs within tighter limits of
intrinsic scatter [0.047,0.079] and slope [-0.094,-0.075].
We find that the CMR universality is robust to changes in sample size
and sampling radii.
Moreover, we show that CMR uniformity extends to $M^{\star}_V+3$ and
we find no marked change in the slope of the $(U-V)$ relation in contrast
with the study of \citet{metcalfe94}.  Last, we find that
CMRs based on colors measured in
apertures containing the same fraction of light are inconsistent with
flat (i.e. zero slope) as claimed by \citet{scodeggio01}.

The CMR results that we present here
provide an useful benchmark for comparisons with CMRs studied in high
redshift clusters using similar rest-frame colors.
The variation in CMR scatter we have observed among four nearby Abell clusters
\citep[the three in our sample plus Coma; BLE92][]{terlevich01} is
consistent with the values reported in the literature at higher redshifts.
This lack of evidence for evolution of the $(U-V)$ CMR scatter suggests,
as the narrow CMR scatter among our three clusters does, that the bulk of
stars in the red-sequence galaxies formed above $z\sim1$.

We use the CMR, spatial, kinematic, and morphological data to identify the
population of galaxies most recently accreted by the clusters.
We first divide each cluster sample into
three populations based on $(U-V)$ color relative to the well-defined CMR.
The colors of cluster galaxies are likely to depend on their time since
arrival \citep{balogh00}; thus, bluer relative color should
correspond to later accretion.  
We establish the existence of significant numbers ($18-23\%$)
of blue and moderately blue galaxies among the luminous ($\ge0.1L^{\star}$)
residents of these Abell clusters.  The blue members have spatial,
kinematic, and morphological properties that are inconsistent with
their red cluster neighbors.  In general, we find bluer galaxies prefer 
the cluster outskirts, avoid the inner regions, have flat and off-center
velocity distributions, and have late-type morphologies.  These properties
provide compelling evidence confirming the more recent
arrival nature of members with very blue colors relative to the
well-defined CMR.  Moreover, the members with intermediate colors are
possibly objects in ``transition'' between more recent arrival VBGs and
long-member RSGs.
The existence of late cluster arrivals provides the best present-day location
to look for evidence of environmental-induced galaxy evolution in cluster 
environments.  To further comment on how the most recent arrival populations fit
into the hierarchical picture of cluster galaxy evolution
requires more detailed scrutiny (profile model fitting with proper
accounting of selection effects) of the blue membership and direct 
comparison with blue field galaxies, all of which we address in Paper 1.

\acknowledgments{}
We are grateful to Frank Valdes for his help with the
early versions of the {\sc mscred} package for reducing Mosaic images.
For useful discussions and correspondence we thank Eric Bell, 
Roelof de Jong, Rob Kennicutt, Gary Schmidt, Luc Simard, 
Ian Smail, Matthias Steinmetz, and
Dennis Zaritsky.
For help with technical aspects of data reduction and photometry,
many kind regards go to Lindsey Davis, Ian Dell'Antonio, 
Rose Finn, Mike Fitzpatrick,
Paul Harding, Mario Hamuy, George Jacoby, Buell Jannuzi, Tod Lauer, Huan Lin,
Mike Meyer, Chien Peng, Cathy Petry, and Greg Rudnick.
We thank also the staff at the KPNO 0.9-meter Telescope for their help and
support during our observing runs.
DHM acknowledge support from the National Aeronautics and Space Administration 
(NASA) under LTSA Grant NAG5-13102.  AIZ is supported by NSF grant
AST-0206084 and NASA LTSA grant NAG5-11108.
Finally DHM thanks the remainder of his thesis committee -- Craig Foltz, Chris
Impey, Ed Olszewski, and Daniel Eisenstein -- for insightful comments that
ultimately improved this paper.  Finally, we thank the anonymous referee
for a thorough and useful review.
This research has made extensive use of NASA's Astrophysical Data
System Abstract Service (ADS) and the astro-ph/ preprint server.

\appendix
\section{CMR Dependence on Color Aperture Choice}
\label{apertures}

We test how the choice of color aperture (fixed light, identical light 
fractions) affects the measurement of CMR slope, scatter, and zero point.
The use of circular apertures with fixed metric sizes
to measure the colors of galaxies
does not take into account their different physical sizes.
BLE92 noted that CMR slopes will differ between those based on fixed metric and
on relative size apertures.  \citet{scodeggio01} have taken this one step 
further, arguing that fixed aperture colors introduce a bias in the CMR because
early-type galaxies have radial color gradients, with redder centers 
compared to outskirts \citep[see e.g.][]{franx89,peletier90}.
Therefore, galaxies of different intrinsic size (which is related to 
luminosity) will have different total fractions of their light measured
within fixed apertures, which may result in a spurious correlation between
galaxy brightness and color.
\citet{scodeggio01} found that the CMR slope of Coma is significantly
diminished to a value consistent with zero, and the intrinsic scatter
is much larger, when the BLE92 sample is re-analyzed using half-light
apertures.  Similarly, \citet{bernardi03} found that $(g-r)$ is independent 
of luminosity for 9000 low-redshift field ellipticals when the total
(model-dependent) galaxy light is used to define color.
Clearly, measuring colors in apertures containing
the same fraction of light is important when attempting to establish
the universality of nearby cluster CMRs.  In particular, we test whether
the use of variable size apertures based on the extent of each galaxy light 
profile will produce flattened CMRs.

Besides the use of fixed metric apertures,
we measure the $(U-V)$ color of all cluster galaxies using two additional
circular apertures with radii given by factors of $1\times$ and $2\times$ the
half-light radius $r_{\rm hl}$.  Our 2D surface brightness fitting
(\S~\ref{stargal}) provides PSF-convolved \rhl measurements for each galaxy.
The use of aperture sizes based on model fit radii introduces an error
in color from uncertainties in \rhl.  We estimate this color error and find
it to be negligible in 98\% of all cluster galaxies; in the remaining 2\% 
this error is roughly $0.01-0.02$ mag.
We note that using such apertures is useful only if the galaxies are resolved.
We find 95\% (A85), 100\% (A496),
and 89\% (A754) of cluster members to be fully resolved such that twice
their half-light size exceeds the common large PSF value we use to
degrade the images for color measurements.  We require the small fraction of
unresolved galaxies to have a limiting aperture diameter
equivalent to the large PSF size.  We measure colors within the half-light
related apertures following the procedure we outline in \S~\ref{colors},
and we present these measurements electronically (see Table~6).

\begin{figure*}[hp]
\center{\includegraphics[scale=.85, angle=0]{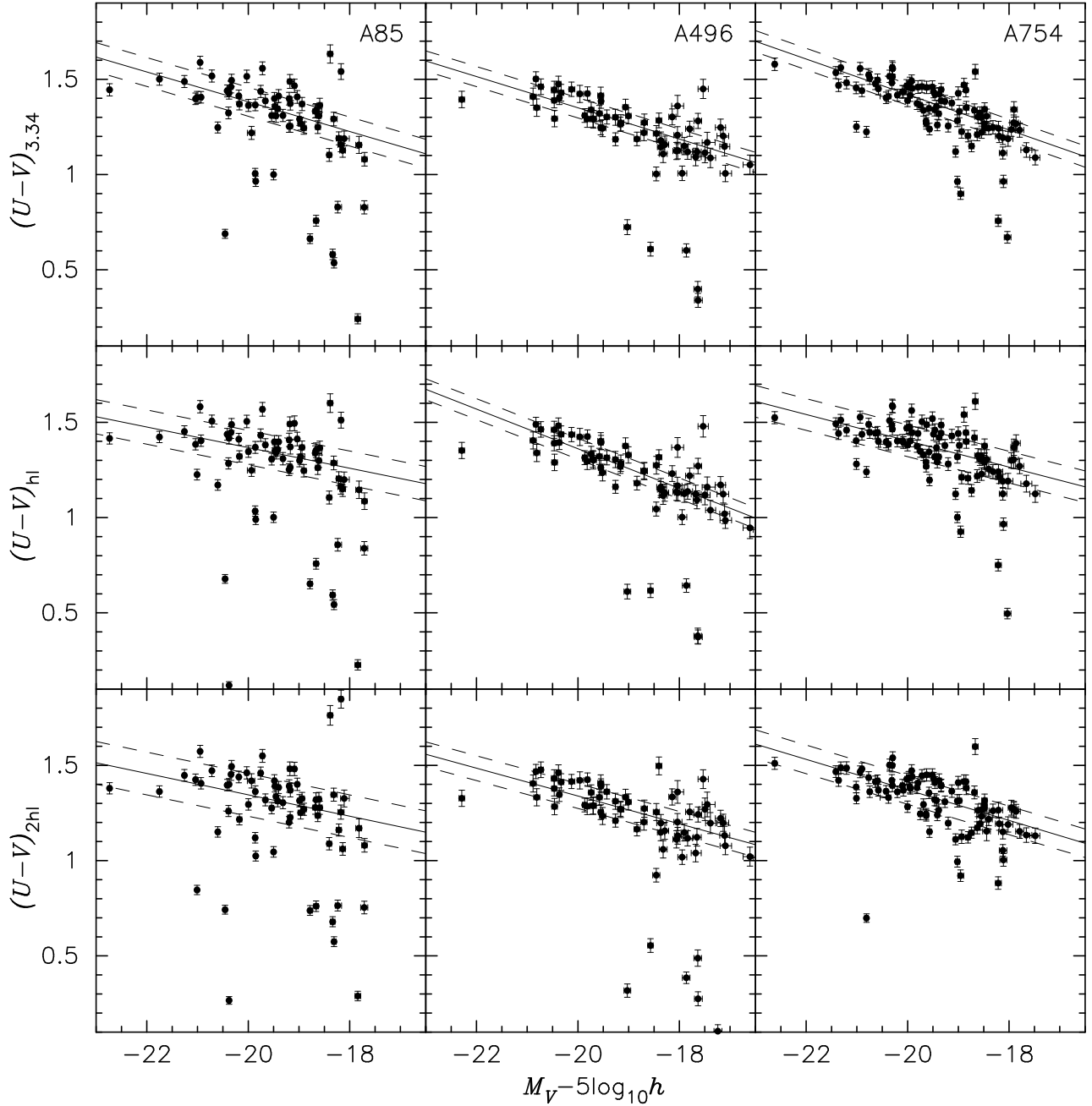}}
\caption
{Comparison of CMRs based on $(U-V)$ colors using three different aperture
diameters: fixed physical size of $3.34$~\hkpc (top panels),
half-light (middle panels), and twice half-light (bottom panels).  
In each panel we plot the
C-M data for cluster galaxies within a projected cluster-centric radius of 
$1/3R_{200}$.  The best-fit CMR (solid line) and its $\pm1\sigma_{\rm CMR}$ 
intrinsic dispersion (dashed lines) from our maximum-likelihood fit to the
data are shown for each aperture and each cluster (see also
Table~8).
\label{CMRcomp}}
\vspace{-0.2cm}
\end{figure*}

We apply our CMR
analysis to members selected within our default sampling radius of
$R_{\rm p}=1/3R_{200}$ and present the results for all aperture color
measurements in Table~8 (see rows 2-4, 8-10, and 15-18).  
\rp is the projected cluster-centric 
radius (relative to the brightest cluster galaxy), 
and $R_{200}$ is a measure of the virial radius.
We do not find dramatic differences between CMR parameters
based on fixed or variable
aperture colors within a given cluster.  In general, the results for
the three different apertures are within $2-3\sigma$ of each other.
We illustrate the similarity in Figure \ref{CMRcomp} where
we plot the C-M data based on each aperture for cluster galaxies
within $1/3R_{200}$.  We note that the top panels of Figure \ref{CMRcomp}
show the uniformity of the three cluster CMRs when using fixed metric
apertures (\S \ref{universality}).  
We find a $2-3\sigma$ (slope measurement error) flattening of the 
CMR slope when using
colors defined in large, roughly total light apertures (i.e. $2r_{\rm hl}$,
bottom panels), compared with the fixed metric aperture results (top
panels).  The slope flattening is greatest for A85 and minimal for A754.  
Nevertheless, even
twice half-light apertures {\it do not} remove the CMR slope in any of 
the three clusters as claimed by \citet{scodeggio01}.  

We note that the
only case where the three cluster CMRs are statistically different
is found when using half-light colors (middle panels of Figure \ref{CMRcomp}
and rows 3, 9, and 16 of Table~8).  Here the slope of A496
is significantly steeper than those of A85 and A754, which are equivalent.
A496 is a smaller and less massive cluster than the other two,
and it is the closest of the three.  It is not clear if these
factors contribute to the observed half-light CMR differences; nevertheless,
these differences represent the maximum range of CMR slopes that we report.

A small fraction of cluster galaxies in our sample are not 
resolved (i.e. half-light size less than seeing size, see
\S \ref{colors}).  Unresolved galaxies have colors derived from apertures
fixed to the seeing size.  Therefore, we check whether the fraction of
unresolved galaxies included in the CMR analysis could artificially
preserve the statistically significant slope we find with half-light
aperture colors.  We repeat our CMR analysis on the resolved
members towards the center of A754, which has the largest fraction of
unresolved galaxies.  This selection reduces the number of
galaxies defining the CMR by $\sim13\%$ (11/82), yet as shown in
Table~8, the best-fit parameters for the smaller sample of 
71 resolved red members is an exact match to the CMR including the 11
additional unresolved galaxies.
The number of such objects within $1/3R_{200}$ of A85 and A496 
are 3 and 0, respectively.  Therefore, we assume that
their removal will also have no effect on the half-light aperture CMR.
We conclude that the seeing of our observations has no
effect on our comparisons of fixed and variable size apertures.

Moreover, increased aperture size produces a small
systematic blueward shift of
the CMR.  The difference in CMR zero point from
$1r_{\rm hl}$ to $2r_{\rm hl}$ aperture colors is $\Delta(U-V)_0=-0.03$ (A85),
$\Delta(U-V)_0=-0.02$ (A496), and $\Delta(U-V)_0=-0.03$ (A754), which
corresponds to $\sim2\sigma$.  These
differences are roughly $3\sigma$, and thus a real
consequence of the color gradients in the presumably early-type
(red) galaxies defining the CMR.

Finally, the trends in intrinsic CMR scatter change with color aperture
choice are less well defined among
the clusters.  Increasing the color aperture from $1r_{\rm hl}$ to $2r_{\rm hl}$
results in $0.02$ and $0.01$ mag increases in \cmrsig for A85 and A496,
respectively.  Yet, the CMR from larger aperture colors in A754 is
roughly 0.01 mag tighter than with the smaller variable aperture.
Statistically, these \cmrsig differences are unimportant 
($1-2\sigma$, scatter measurement error).
The only significant ($>3\sigma$) change occurs for A85 going from fixed-size
to $2r_{\rm hl}$ apertures.  For all three clusters
we find the minimum intrinsic scatter when using fixed aperture colors.

To summarize, measuring color with galactic half-light radii instead of
fixed metric apertures increases the CMR scatter by at most 33\% and
nearly quadruples the variation in CMR slope among the sample
clusters.

\onecolumn

\begin{deluxetable}{lcccccccccc}
\tablewidth{0pt}
\tablenum{1}
\tabletypesize{\small}
\tablecolumns{11}
\tablecaption{Cluster Properties}
\label{clus_props}
\tablehead{\colhead{Cluster} & \colhead{$\alpha_{2000}$} & \colhead{$\delta_{2000}$} & \colhead{$l$} & \colhead{$b$} & \colhead{$z$} & \colhead{R} & \colhead{$R_{\rm vir}$} & \colhead{$R_{\rm X}$} & \colhead{$M_{\rm cv}$} & \colhead{$M_{\rm X}$} \\
\colhead{(1)} & \colhead{(2)} & \colhead{(3)} & \colhead{(4)} & \colhead{(5)} &
\colhead{(6)} & \colhead{(7)} & \colhead{(8)} & \colhead{(9)} & \colhead{(10)} & \colhead{(11)} }
\startdata
A85  & 00:41:50.5 & -09:18:11.6 & 116.24 & -72.03 & 0.055 & 1 & 1.94 & 0.7-1.5 & 9.88 & 1.3-6.3 \\
A496 & 04:33:37.8 & -13:15:43.5 & 210.59 & -36.49 & 0.033 & 1 & 1.37 & 0.3-0.6 & 3.20 & 0.9-1.9 \\
A754 & 09:08:32.4 & -09:37:46.5 & 187.67 & +24.71 & 0.055 & 2 & 1.32 & 5.0     & 4.23 & 0.6    \\
\enddata
\tablecomments{Abell catalog designation (1).  Epoch J2000.0 celestial
(2,3) and
Galactic (4,5 in degrees) coordinates of the image center coincident
with the brightest cluster galaxy (a cD galaxy for clusters A85 and A496; a D
class galaxy in the case of A754).  The mean
cluster redshift (6).  The Abell richness (7), which
is defined by \citet{abell58} as the number of galaxies $N_{\rm gal}$ 
in a cluster with $m\leq m_3 +2$, where $m_3$ is the third brightest member.
The richness values reported here correspond to
$50\leq N_{\rm gal} \leq 79$ (R=1), and
$80\leq N_{\rm gal} \leq 129$ (R=2).
From \citet{girardi98}, 
the virial (8) and X-ray derived (9) radii
in units of \hmpc, and the virial (10) and X-ray (11) masses
in units of $10^{14} h^{-1}$~M$\sun$ within corresponding radii. }
\end{deluxetable}
\begin{deluxetable}{llcccccccl}
\tablewidth{0pt}
\tablenum{2}
\tabletypesize{\small}
\tablecolumns{10}
\tablecaption{Cluster Observations}
\label{clus_obs}
\tablehead{\colhead{Cluster} & \colhead{Date} & \colhead{Filter} & \colhead{Camera} & \colhead{$t_{\rm exp}$} & \colhead{Seeing} & \colhead{$\sec{z}$} & \colhead{FOV} & \colhead{scale} & \colhead{Conditions} \\
\colhead{(1)} & \colhead{(2)} & \colhead{(3)} & \colhead{(4)} & \colhead{(5)} & \colhead{(6)} & \colhead{(7)} & \colhead{(8)} & \colhead{(9)} & \colhead{(10)} }
\startdata
A85  & Nov. 28, 1997 & $V$ & MOS1 & $5\times600$         & 1.2 & 1.33-1.36 & 2.7 & 0.32 & photometric \\
          & Jan. 04, 2001 & $U$ & MOS2 & $5\times600$         & 1.4 & 1.41-1.58& & & photometric \\
A496 & Nov. 25, 1997 & $V$ & MOS1 & $5\times600$         & 1.3 & 1.42-1.47 & 1.6 & 0.19 & non-photometric \\
          & Nov. 27, 1997 & $V$ & MOS1 & $1\times300$ & 1.2 &    1.43& & & photometric \\
          & Jan. 04, 2001 & $U$ & MOS2 & $5\times720$         & 1.3 & 1.42-1.44& & & photometric \\
A754 & Nov. 28, 1997 & $V$ & MOS1 & $5\times480$         & 1.1 & 1.34-1.36 & 2.7 & 0.31 & photometric \\
          & Nov. 28, 1997 & $V$ & MOS1 & $1\times300$ & 1.1 &    1.34& & & photometric \\
          & Jan. 3, 2000  & $U$ & MOS2 & $3\times720$         & 2.4 & 1.37-1.43& & & poor seeing \\
          & Jan. 4, 2000 & $U$ & MOS2 & $5\times720$          & 2.0 & 1.34-1.39& & & photometric \\
\enddata
\tablecomments{For each cluster observation we give
the date (2), filter (3), and detector (4).  The exposure time (5) is
given by the number of exposures multiplied by the duration of each
integration.  Single exposures were obtained for calibration during photometric
conditions.  The average seeing FWHM (6) in arcseconds and the range
of airmass (7) given as $\sec z$ for a set of exposures.  For each cluster
we provide the physical scale of the square-degree diameter field of view (8; FOV)
in units of $h^{-1}$~Mpc,
and the physical pixel scale (9) in units of $h^{-1}$~kpc~pix$^{-1}$.
We note the conditions during observations in (10).}
\end{deluxetable}
\begin{deluxetable}{lcccccc}
\tablewidth{0pt}
\tablenum{3}
\tabletypesize{\small}
\tablecolumns{7}
\tablecaption{Final Cluster Image Properties}
\label{im_props}
\tablehead{\colhead{Cluster} & \colhead{Filter} & \colhead{$\left<{\rm bkg}\right>$} & \colhead{$\sigma_{\rm bkg}$} & \colhead{$\tilde{t}_{\rm exp}$} & \colhead{$\tilde{X}$} & \colhead{$zp_{\rm eff}$} \\
\colhead{(1)} & \colhead{(2)} & \colhead{(3)} & \colhead{(4)} & \colhead{(5)} & \colhead{(6)} & \colhead{(7)} }
\startdata
A85  & V & 45.52 & 1.61 & 600 & 1.33 & 26.894 \\
     & U & 17.75 & 1.88 & 600 & 1.41 & 26.629 \\
A496 & V & 37.67 & 1.25 & 300 & 1.42 & 26.162 \\
     & U & 18.17 & 1.84 & 720 & 1.44 & 26.814 \\
A754 & V & 31.09 & 1.36 & 480 & 1.34 & 26.651 \\
     & U & 22.74 & 1.56 & 720 & 1.34 & 26.857 \\
\enddata
\tablecomments{Properties of each fully reduced cluster image are passband (2),
average background level in adu (3), RMS pixel-to-pixel noise in adu (4),
effective exposure time in seconds (5), effective airmass (6), and
effective zero point (7).
}
\end{deluxetable}
\begin{deluxetable}{lccccccl}
\tablewidth{0pt}
\tablenum{4}
\tabletypesize{\scriptsize}
\tablecolumns{8}
\tablecaption{Photometric Transformation Equation Coefficients}
\label{te_coeffs}
\tablehead{\colhead{Date} & \colhead{Filter} & \colhead{Scope/Detector} & \colhead{$N_{\rm stars}$} & \colhead{fit RMS} & \colhead{$zp$} & \colhead{$\alpha$} & \colhead{$\beta$} \\
\colhead{(1)} & \colhead{(2)} & \colhead{(3)} & \colhead{(4)} & \colhead{(5)} & \colhead{(6)} & \colhead{(7)} & \colhead{(8)} }
\startdata
Nov. 27, 1997 & V & 0.9-meter/MOS1 & 44 & 0.029 & $20.161\pm0.012$ & $-0.135\pm0.006$ & $-0.003\pm0.008$ \\
Nov. 28, 1997 & V & 0.9-meter/MOS1 & 34 & 0.029 & $20.063\pm0.026$ & $-0.086\pm0.014$ & $0.009\pm0.008$ \\
Jan. 4, 2000 & U & 0.9-meter/MOS2 & 112 & 0.058 & $20.287\pm0.028$ & $-0.428\pm0.016$ & $-0.018\pm0.012$ \\
\enddata
\tablecomments{For each night that photometric standards were
observed we give the date (1),
filter (2), and scope plus detector system (3).  The best-fit solution
to the transformation equation for photometric calibration was achieved
using the given number of unique standard stars (4), with an RMS in
magnitudes (5), and with best-fit coefficients and uncertainties given
for the zero point (6), airmass (7), and color (8) terms.
}
\end{deluxetable}
\begin{deluxetable}{lcccccc}
\tablewidth{0pt}
\tablenum{5}
\tabletypesize{\small}
\tablecolumns{7}
\tablecaption{Cluster Image Source Catalogs}
\label{SourceCats}
\tablehead{\colhead{Cluster} & \colhead{Image} & \colhead{$N_{\rm good}$} & \colhead{$m_{\rm min}$} & \colhead{$N_{\rm min}$} & \colhead{$N_{\rm ext}$} & \colhead{$N_{\rm gal}^{U,V}$} \\
\colhead{(1)} & \colhead{(2)} & \colhead{(3)} & \colhead{(4)} & \colhead{(5)} & \colhead{(6)} & \colhead{(7)} }
\startdata
A85  & $V$   & 4128 & 20.5 & 2786 & 1448 & \\
     & $U$   & 1574 & 20.7 & 1453 & & \\
     & total & & & & & 455 \\
A496 & $V$   & 3301 & 20.5 & 3063 & 730 & \\
     & $U$   & 2183 & 21.0 & 2043 & & \\
     & total & & & & & 315 \\
A754 & $V$   & 8786 & 21.0 & 7578 & 2142 & \\
     & $U$   & 3924 & 20.5 & 3451 & & \\
     & total & & & & & 547 \\
\enddata
\tablecomments{Breakdown of cluster source catalogs.  The image catalog
passband or `total', which denotes the final $U$ and $V$ combined catalog,
is designated in (2).
Total number of `good' sources (3) after the removal of saturated, bad, and
edge objects.  Empirical magnitude limit (4) and corresponding number of
sources brighter than $m_{\rm min}$ (5).  Number of `extended' sources
in the $V$-band image (6), and final quantity of $V$-band galaxies with
$U$ counterparts (7).
}
\end{deluxetable}
\begin{deluxetable}{lcccccccc}
\tablewidth{0pt}
\tablenum{6}
\tabletypesize{\scriptsize}
\tablecolumns{9}
\tablecaption{Example of the Catalog of Cluster Galaxy Photometry}
\label{MemPhot}
\tablehead{\colhead{Galaxy} & \colhead{$z$} & \colhead{$M_V-5\log_{10}h$} & \colhead{$(U-V)_{3.34}$} & \colhead{$(U-V)_{\rm hl}$} & \colhead{$(U-V)_{\rm 2hl}$} & \colhead{$E(B-V)$} & \colhead{$k_V$} & \colhead{$k_{(U-V)}$} \\
\colhead{(1)} & \colhead{(2)} & \colhead{(3)} & \colhead{(4)} & \colhead{(5)} & \colhead{(6)} & \colhead{(7)} & \colhead{(8)} & \colhead{(9)}}
\startdata
A85.004150$-$091811 & 0.055 & $-22.73\pm0.03$ & $1.44\pm0.03$ & $1.42\pm0.03$ & $1.38\pm0.03$ & 0.04 & -0.10 & -0.08 \\
A85.003959$-$092604 & 0.057 & $-18.61\pm0.05$ & $0.65\pm0.03$ & $0.67\pm0.03$ & $0.83\pm0.03$ & 0.03 & -0.10 & -0.08 \\
A85.004005$-$090303 & 0.056 & $-20.35\pm0.03$ & $1.66\pm0.03$ & $1.61\pm0.03$ & $1.58\pm0.03$ & 0.03 & -0.10 & -0.08 \\
A85.004013$-$085957 & 0.055 & $-19.26\pm0.04$ & $1.46\pm0.03$ & $1.46\pm0.03$ & $1.37\pm0.03$ & 0.03 & -0.10 & -0.08 \\
A85.004018$-$085631 & 0.052 & $-18.33\pm0.08$ & $1.30\pm0.04$ & $1.30\pm0.04$ & $1.26\pm0.03$ & 0.03 & -0.09 & -0.07 \\
A496.043256$-$133639 & 0.038 & $-20.63\pm0.07$ & $0.65\pm0.05$ & $0.67\pm0.05$ & $0.70\pm0.05$ & 0.18 & -0.07 & -0.06 \\
A496.043258$-$131218 & 0.036 & $-19.25\pm0.05$ & $1.19\pm0.03$ & $1.16\pm0.03$ & $1.21\pm0.03$ & 0.08 & -0.07 & -0.05 \\
A496.043259$-$131820 & 0.034 & $-17.83\pm0.10$ & $1.12\pm0.04$ & $1.14\pm0.04$ & $1.12\pm0.04$ & 0.10 & -0.06 & -0.05 \\
A496.043300$-$131600 & 0.037 & $-20.12\pm0.06$ & $1.45\pm0.04$ & $1.44\pm0.04$ & $1.41\pm0.04$ & 0.09 & -0.07 & -0.05 \\
A496.043300$-$134232 & 0.035 & $-19.39\pm0.09$ & $1.04\pm0.06$ & $0.90\pm0.06$ & $0.80\pm0.06$ & 0.23 & -0.06 & -0.05 \\
A754.090645$-$093424 & 0.060 & $-17.74\pm0.08$ & $0.22\pm0.03$ & $0.21\pm0.03$ & $0.21\pm0.03$ & 0.07 & -0.11 & -0.08 \\
A754.090803$-$100214 & 0.054 & $-20.09\pm0.05$ & $1.65\pm0.04$ & $1.62\pm0.04$ & $1.61\pm0.04$ & 0.08 & -0.10 & -0.08 \\
A754.090806$-$092600 & 0.054 & $-18.94\pm0.05$ & $0.77\pm0.03$ & $0.79\pm0.03$ & $0.74\pm0.03$ & 0.06 & -0.10 & -0.08 \\
A754.090806$-$092713 & 0.051 & $-19.02\pm0.04$ & $0.54\pm0.02$ & $0.58\pm0.02$ & $0.58\pm0.02$ & 0.06 & -0.09 & -0.07 \\
A754.090807$-$100410 & 0.050 & $-18.07\pm0.08$ & $1.38\pm0.04$ & $1.41\pm0.05$ & $1.38\pm0.04$ & 0.09 & -0.09 & -0.07 \\
\enddata
\tablecomments{The full catalog is available electronically.
Galaxy identification based on cluster designation and
J2000.0 celestial coordinates (1).  Rest-frame $V$-band total absolute 
magnitude (2).  Rest-frame $(U-V)$ colors measured within three 
circular apertures: fixed size corresponding to $3.34$~\hkpc at distance
to parent cluster (3); variable sizes (see 
Appendix \ref{apertures}) corresponding to half-light (4) and
twice half-light (5) galaxy diameters.  The photometry is calibrated to 
Johnson $U$ and $V$ magnitudes on the \citet{landolt92} system, and each
measurement includes formal random errors.
}
\end{deluxetable}
\begin{deluxetable}{lccccccccc}
\tablewidth{0pt}
\tablenum{7}
\tabletypesize{\small}
\tablecolumns{10}
\tablecaption{Cluster Membership Results}
\label{Members}
\tablehead{\colhead{Cluster} & \colhead{$\left<cz\right>_{\rm clus}$} & \colhead{$\sigma_{\rm clus}$} & \colhead{$DM$} & \colhead{$R_{200}$} & \colhead{$N_z$} & \colhead{$N_{\rm mem}$} & \colhead{$N_{\rm non}$} & \colhead{$N_{\rm miss}$} & \colhead{$\theta_{\rm sep}$}  \\
\colhead{(1)} & \colhead{(2)} & \colhead{(3)} & \colhead{(4)} & \colhead{(5)} & \colhead{(6)} & \colhead{(7)} & \colhead{(8)} & \colhead{(9)} & \colhead{(10)} }
\startdata
A85  & $16607\pm60$ & $993\pm53$ & 36.184 & 1.67 & 331 & 180 (0.4/0.5/0.1) & 80 & 71 (27) & 1.15\\
A496 &  $9910\pm48$ & $728\pm36$ & 35.030 & 1.24 & 225 & 146 (0.2/0.6/0.2) & 26 & 53 (33) & 0.90\\
A754 & $16369\pm47$ & $953\pm40$ & 36.158 & 1.61 & 415 & 311 (0.3/0.6/0.1) & 50 & 54 (24) & 1.66\\
\enddata
\tablecomments{For each cluster (1), the mean recessional velocity (2) and
internal velocity dispersion (3), in units of km~$s^{-1}$, are given from the
spectroscopic survey of \citet{christlein03}.  The resultant cosmological
distance modulus assuming $h=1$ (4), and $R_{200}$ (5) in units of $h^{-1}$~Mpc
in our adopted cosmology,
which is an estimate of the cluster virial radius defined to encompass a density
that is 200 times the critical density at a given redshift \citep{finn04}.
The total number of redshifts within our images (6), which is divided into
the following quantities: members meeting the membership criteria
(see text) with $U,V$ image source counterparts (7); non-members with 
$U,V$ image sources
(8); and spectroscopic sources without image matches (9), with the number of
unmatched member redshifts given in parentheses.
The average coordinate separations (10)
between the imaging and spectroscopic matchups.  Within the parentheses of 
column (7) we give the relative fractions of bright member galaxies 
($\ge0.1L^{\star}$) split into three basic visual morphology types (E/S0/S+Irr).
}
\end{deluxetable}
\begin{deluxetable}{lccccccc}
\tablewidth{0pt}
\tablenum{8}
\tabletypesize{\small}
\tablecolumns{8}
\tablecaption{Best-Fit CMR Analysis Results}
\label{CMRresults}
\tablehead{\colhead{Cluster} & \colhead{Ap} & \colhead{$R_{\rm p}/R_{200}$} & \colhead{$N_{\rm fit}$} & \colhead{$(U-V)_0$} & \colhead{\cmrsig} & \colhead{$d(U-V)/dM_V$} & \colhead{$\mathcal{L}_{\rm max}$} \\
\colhead{(1)} & \colhead{(2)} & \colhead{(3)} & \colhead{(4)} & \colhead{(5)} & \colhead{(6)} & \colhead{(7)} & \colhead{(8)}}
\startdata
  A85 & 1 & $0.30(0.50)$ &  48/ 62 & $1.38\pm0.01$ & $0.079\pm0.010$ & $-0.075\pm0.012$ & 49.8 \\
  A85 & 1 & $0.33(0.56)$ &  53/ 67 & $1.38\pm0.01$ & $0.076\pm0.008$ & $-0.078\pm0.010$ & 55.9 \\
  A85 & 2 & $0.33(0.56)$ &  54/ 67 & $1.37\pm0.01$ & $0.090\pm0.011$ & $-0.054\pm0.012$ & 48.4 \\
  A85 & 3 & $0.33(0.56)$ &  55/ 67 & $1.34\pm0.02$ & $0.112\pm0.010$ & $-0.056\pm0.014$ & 40.8 \\
  A85 & 1 & $0.37(0.60)$ &  60/ 78 & $1.39\pm0.01$ & $0.074\pm0.009$ & $-0.077\pm0.010$ & 65.7 \\
 A496 & 1 & $0.25(0.31)$ &  45/ 50 & $1.35\pm0.01$ & $0.060\pm0.009$ & $-0.077\pm0.010$ & 53.2 \\
 A496 & 1 & $0.30(0.37)$ &  50/ 60 & $1.35\pm0.01$ & $0.047\pm0.007$ & $-0.087\pm0.008$ & 69.3 \\
 A496 & 1 & $0.33(0.41)$ &  60/ 71 & $1.35\pm0.01$ & $0.053\pm0.009$ & $-0.082\pm0.008$ & 77.4 \\
 A496 & 2 & $0.33(0.41)$ &  60/ 71 & $1.36\pm0.01$ & $0.055\pm0.009$ & $-0.104\pm0.009$ & 77.1 \\
 A496 & 3 & $0.33(0.41)$ &  60/ 71 & $1.34\pm0.01$ & $0.065\pm0.009$ & $-0.073\pm0.009$ & 69.1 \\
 A496 & 1 & $0.37(0.46)$ &  70/ 80 & $1.35\pm0.01$ & $0.053\pm0.008$ & $-0.081\pm0.007$ & 87.4 \\
 A496 & 1 & $0.50(0.60)$ &  91/111 & $1.35\pm0.01$ & $0.055\pm0.007$ & $-0.089\pm0.010$ & 111.7 \\
 A754 & 1 & $0.20(0.32)$ &  52/ 58 & $1.42\pm0.01$ & $0.055\pm0.008$ & $-0.089\pm0.008$ & 68.9 \\
 A754 & 1 & $0.30(0.48)$ &  75/ 83 & $1.42\pm0.01$ & $0.050\pm0.009$ & $-0.093\pm0.009$ & 105.4 \\
 A754 & 1 & $0.33(0.54)$ &  86/ 95 & $1.42\pm0.01$ & $0.056\pm0.007$ & $-0.093\pm0.007$ & 111.0 \\
 A754 & 2 & $0.33(0.54)$ &  89/ 95 & $1.40\pm0.01$ & $0.083\pm0.009$ & $-0.069\pm0.010$ & 89.0 \\
 A754 & $2^{a}$ & $0.33(0.54)$ & 77/ 82 & $1.40\pm0.01$ & $0.084\pm0.009$ & $-0.071\pm0.011$ & 76.5 \\
 A754 & 3 & $0.33(0.54)$ &  89/ 95 & $1.37\pm0.01$ & $0.075\pm0.009$ & $-0.080\pm0.008$ & 95.9 \\
 A754 & 1 & $0.37(0.60)$ &  95/107 & $1.43\pm0.01$ & $0.061\pm0.008$ & $-0.094\pm0.008$ & 116.7 \\
\enddata
\tablecomments{Cluster designation (1).  Aperture type (2) for 
$(U-V)$ color measurement:
1 = fixed size corresponding to $3.34$~\hkpc at distance to
cluster; 2 = variable size corresponding to half-light galaxy diameter;
3 = variable size corresponding to twice half-light galaxy diameter.
The projected cluster-centric radius in terms of $R_{200}$ is given in
(3), and in units of \hmpc in parentheses.
For comparison, the Coma cluster has $R_{200}=1.76 h^{-1}$~Mpc based
on $z=0.022$ and $\sigma_{\rm clus}=1008$ km~$s^{-1}$ from \citet{struble99}.
Within projected radius
$R_{\rm p}$, the number of galaxies used in the final fit (after
outlier clipping, see text) relative to the total number are given in (4).
The best-fit CMR parameters from our maximum likelihood analysis
are zero point (5), intrinsic scatter (6), and slope (7), and there
associated uncertainties.  The maximum likelihood for each fit is
given in (8).
}
\tablenotetext{a}{Results based on C-M data for resolved galaxies only.}
\end{deluxetable}
\begin{deluxetable}{lccccc}
\tablewidth{0pt}
\tablenum{9}
\tabletypesize{\small}
\tablecolumns{6}
\label{CMpops}
\tablecaption{Color-magnitude Galaxy Population Breakdown}
\tablehead{\colhead{Cluster} & \colhead{$L/L^{\star}$} & \colhead{$N_{\rm mem}$} & \colhead{$N_{\rm RSG}$} & \colhead{$N_{\rm IBG}$} & \colhead{$N_{\rm VBG}$} \\
\colhead{(1)} & \colhead{(2)} & \colhead{(3)} & \colhead{(4)} & \colhead{(5)} & \colhead{(6)} }
\startdata
A85  & $\ge0.1$ & 164 & 126 (56,68,2) & 14 (4,7,3) & 24 (5,6,13) \\
     & $<0.1$   &  16 &   4 (4,0,0)   &  3 (1,1,1) &  9 (3,2,4) \\
A496 & $\ge0.1$ & 101 &  83 (18,56,9) & 6 (0,5,1) & 12 (1,1,10) \\
     & $<0.1$   &  45 &  29 (14,5,10) & 1 (1,0,0) & 15 (0,0,15) \\
A754 & $\ge0.1$ & 281 & 227 (68,146,13) & 35 (9,18,8) & 19 (2,4,13) \\
     & $<0.1$   &  30 &  21 (17,3,1) & 4 (3,0,1) & 5 (1,0,4) \\
\enddata
\tablecomments{Breakdown of color-based populations for each cluster divided
into two subsamples: brighter than $0.1L^{\star}$ ($M_V=-18.1+5\log_{10}h$)
and fainter.  For each luminosity bin (2), we give the total
number of cluster members (3) split into RSG (4), IBG (5), and VBG (6) types.
We give the number per three visual morphology types (E,S0,S+Irr) in the
parentheses in each column.
}
\end{deluxetable}
\begin{deluxetable}{lccccccc}
\tablewidth{0pt}
\tablenum{10}
\tabletypesize{\small}
\tablecolumns{8}
\label{KinTests}
\tablecaption{Statistical Test Results for Cluster Kinematics}
\tablehead{\colhead{Cluster} & \colhead{C-M Pop.} & \colhead{$\left<\Delta v\right>$} & \colhead{$\sigma$} & \colhead{$N$} & \colhead{K-S test} & \colhead{F-test} & \colhead{t-Test} \\
\colhead{(1)} & \colhead{(2)} & \colhead{(3)} & \colhead{(4)} & \colhead{(5)} & \colhead{(6)} & \colhead{(7)} & \colhead{(8)}}
\startdata
A85  & RSG & -102 &  908 & 130 &         &      &      \\ 
     & VBG & +656 & 1260 &  33 & $>99.9\%$ & 98.9\% & 99.8\% \\
     & IBG & +262 & 1210 &  17 &   70.4\%  & 91.6\% & 99.8\% \\
A496 & RSG & +165 &  727 & 112 &         &      &      \\ 
     & VBG & -245 & 1050 &  17 &   99.3\%  & 99.1\% & 93.6\% \\
A754 & RSG &  +21 &  958 & 248 &         &      &      \\ 
     & VBG & -425 &  969 &  24 &   93.9\%  &  0.1\% & 97.0\% \\
     & IBG &   -3 &  929 &  39 &    0.1\%  &  0.1\% &  0.1\% \\
\enddata
\tablecomments{Results of tests for differences between the kinematic 
distributions of cluster RSGs compared with the blue galaxy (VBG and IBG)
populations.  For the color-selected galaxy populations (2) of each cluster, 
we give the mean (3) and standard deviation (4)
of the relative velocity distributions in \kms.  The number of galaxies in
each sample is given in (5).  In (6) we tabulate the K-S test probability
that the
blue galaxy population is not drawn from the same parent sample as the RSGs.
We give the F-test significance for differences between sample 
variances in (7), and likewise the T-test significance for sample mean 
differences in (8).  We note that for cases where the sample variance is
different at the $>90\%$ level, we use the ``unequal variances'' version
of the T-test (see text).
As a result of the small number of IBGs in A496 ($N=7$), we do not test
for differences between the kinematics of this population and the cluster's
RSGs.
}
\end{deluxetable}

\end{document}